\documentclass[njp,11pt]{iopart}
\usepackage{graphicx}
\expandafter\let\csname equation*\endcsname\relax
\expandafter\let\csname endequation*\endcsname\relax
\usepackage{amsmath}
\usepackage{amsfonts}
\usepackage{soul}
\usepackage{dsfont}
\usepackage{hyperref}
\usepackage{amstext}
\usepackage{braket}
\usepackage[caption=false]{subfig}
\usepackage{braket}
\usepackage{color,soul} 
\usepackage{dcolumn}   
\usepackage{bm}        
\usepackage{tabularx,ragged2e,booktabs,caption}
\usepackage{cellspace}
\renewcommand\vec{\boldsymbol}
\usepackage{floatrow} 
\usepackage{hyperref}
\hypersetup{colorlinks=true, citecolor=blue, urlcolor=blue, linkcolor=blue}
\usepackage{color,soul}
\usepackage{dsfont}
\usepackage[dvipsnames]{xcolor}

\usepackage{amssymb,bbm}
\date{\today}
\newcommand{\overbar}[1]{\mkern 1.5mu\overline{\mkern-1.5mu#1\mkern-1.5mu}\mkern 1.5mu}
\usepackage[margin=2.5cm]{geometry}

\renewcommand{\braket}[1]{\left\langle #1\right\rangle}

\renewcommand{\vec}[1]{\mathbf{#1}}
\newcommand{\p}{\partial}

\begin{document}
\title{Hydrogenic entanglement}

\author{Sofia Qvarfort$^{1,2}$, Sougato Bose$^2$, Alessio Serafini$^2$}
\address{$^1$ QOLS, Blackett Laboratory, Imperial College London,  SW7 2AZ London, United Kingdom  \\
$^2$ Department of Physics and Astronomy, University College London, Gower Street, WC1E 6BT London, United Kingdom   }
\ead{sofiaqvarfort@gmail.com}

\begin{abstract}
Is there any entanglement in the simplest ubiquitous bound system? We study the solutions to the time-independent Schr{\"o}dinger equation for a Hydrogenic system and devise two entanglement tests for free and localised states. For free Hydrogenic systems, we compute the Schmidt basis diagonalisation for general energy eigenstates, and for a Hydrogenic system localised to a three-dimensional Gaussian wavepacket, we demonstrate that measuring its second moments is sufficient for detecting entanglement. Our results apply to any system that exhibits Hydrogenic structure. 
\end{abstract}

\section{Introduction}
Are the electron and proton in a Hydrogen atom entangled? In 1926, Erwin Schr{\"o}dinger successfully predicted the spectral energies of Hydrogen by solving the wave-equation for an electron wavefunction in a potential-well created by the positively charged nucleus~\cite{schrodinger1926quantisierung,schrodinger2003collected}. While the Hydrogenic solutions to Schr{\"o}dinger's equation have now been known for almost a century, the question of whether the two subsystems are entangled has hardly been investigated. This is  most likely due to the prevalence of  the Born--Oppenheimer (BO) approximation~\cite{born1927quantentheorie}, which explicitly assumes that the motion of two subsystems with vastly different masses can be treated separately. While it sometimes follows that entanglement is explicitly removed in the process of applying the BO approximation, this is in fact only true if the approximation is exact~\cite{izmaylov2017entanglement}. Indeed, entanglement can be retained as degrees of the approximation is applied, and can even be used as a measure of the validity of the BO approximation~\cite{bouvrie2014entanglement}. 

In this work, we forego  the BO approximation  completely in order to study the entanglement of the exact Hydrogenic solutions. The ubiquity of Hydrogen in physical, chemical, and biological systems makes it is one of the most well-studied physical systems, and irrespective of the question as to whether entanglement -- if present in Hydrogenic systems -- has any applications, it is important to know whether such a basic textbook entity of physics is intrinsically entangled. 

In recent years, the study of quantum information-processing tasks has demonstrated the importance of entanglement to quantum computing~\cite{jozsa2003role}, quantum cryptography~\cite{ekert1991quantum}, and quantum sensing~\cite{degen2017quantum}. A number of fundamental questions, such as whether Nature is described by collapse theories~\cite{bassi2013models}, or whether gravity is a quantum force~\cite{kiefer2007quantum}, are aided by the quantification and detection of entanglement~\cite{belli2016entangling, bose2017spin, marletto2017gravitationally, belenchia2018quantum,marshman2019locality}. The ability to experimentally determine entanglement for Hydrogenic systems could have implications for the application of Hydrogenic atoms, artificial atoms~\cite{ashoori1996electrons,fujisawa2002allowed}, exiton states~\cite{wannier1937structure, mott1938conduction}, or possibly even mesoscopic systems, such as levitated nanospheres~\cite{millen2015cavity},  which could be engineered to interact via a central potential~\cite{qvarfort2018mesoscopic}. In fact, the entanglement of similar systems, such as electron-electron entanglement in Helium and Helium-like atoms~\cite{dehesa2011quantum,benenti2013entanglement, lin2013spatial}, and proton-proton entanglement in Hydrogen molecules~\cite{ding2020correlation}, has been recently explored.

\begin{figure*}
\subfloat[ \label{fig:free}]{
  \includegraphics[width=0.5\linewidth, trim = 10mm -10mm 10mm 10mm]{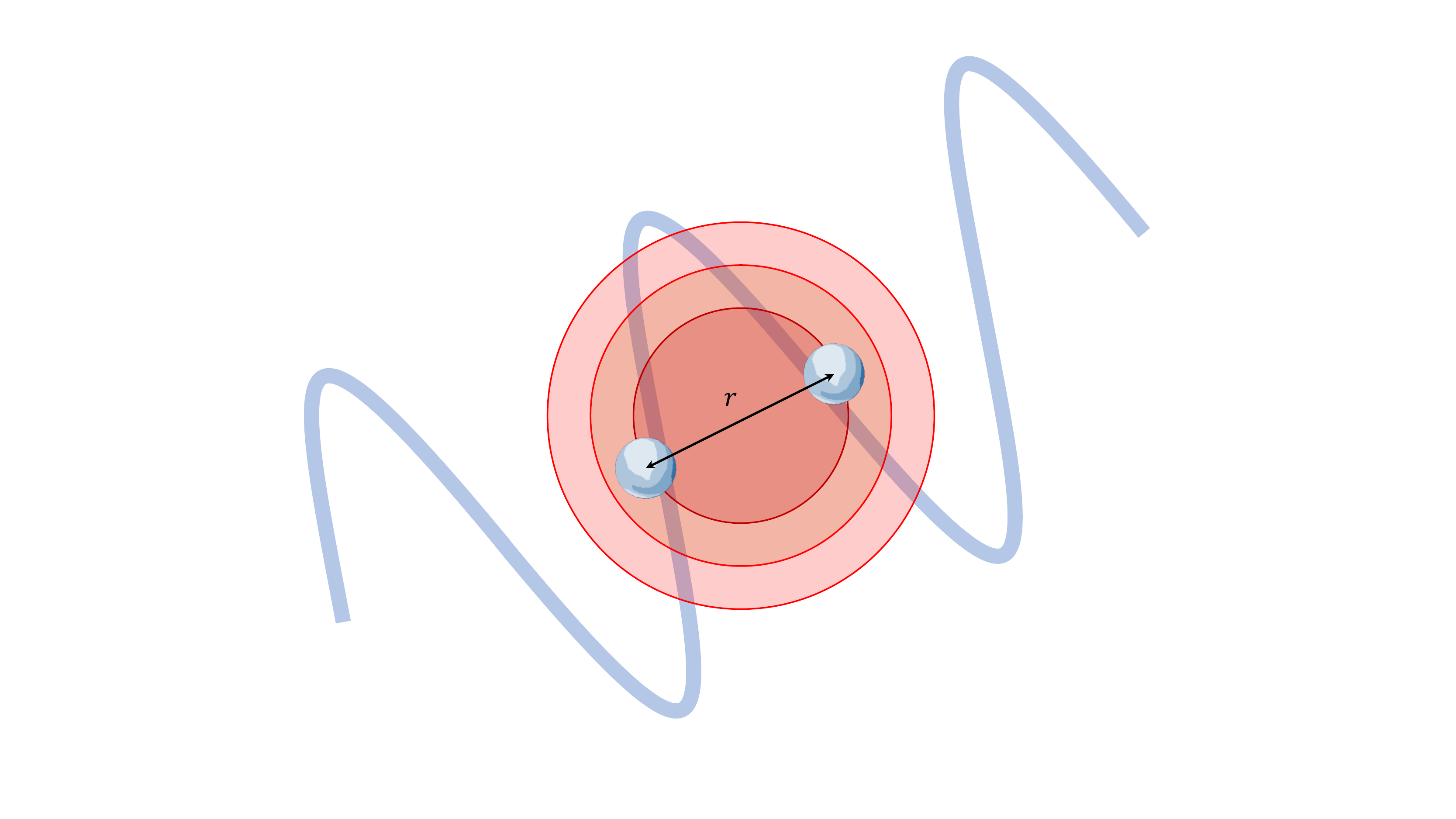}
}  
\subfloat[ \label{fig:Gaussian}]{
  \includegraphics[width=0.5\linewidth, trim = 10mm 0mm 10mm 0mm]{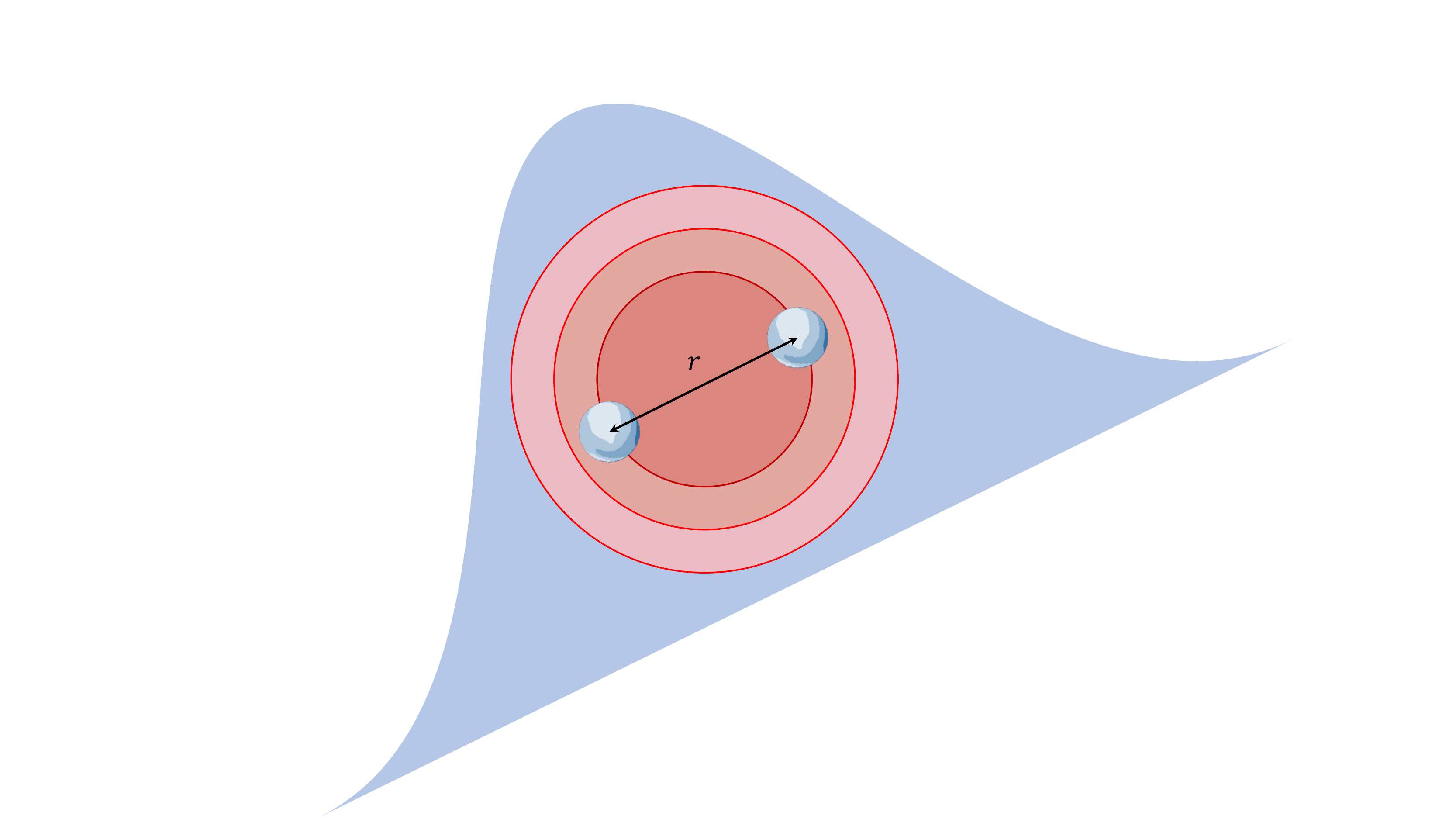}
}
\caption{\small Sketch of a bipartite Hydrogenic system. The relative degree-of-freedom wavepacket takes the Hydrogenic solutions and is parametrised by the variable $r = |\vec{r}_1 - \vec{r}_2|$, where $\vec{r}_1$ and $\vec{r}_2$ are position vectors for the two subsystems.  The centre-of-mass wavepacket is either \textbf{(a)}  a free wavepacket, or \textbf{(b)} a localised three-dimensional Gaussian wavepacket.}
\label{fig:systems}
\end{figure*}

A rudimentary step in certifying the entanglement in Hydrogenic systems has been taken by Tommasini \textit{et al}.~\cite{tommasini1998hydrogen},  who showed that entanglement in a \textit{free} Hydrogenic ground state system can be verified by diagonalising the state through the Fourier transform and computing its Schmidt coefficients. In the laboratory, however, the assumption that the Hydrogenic system is free is no longer accurate.  Rather, any prepared state will be localised to a finite spatial volume, and thus can no longer be diagonalised by the Fourier transform. We must therefore devise an appropriate entanglement test that holds even for localised states,  which is what we set out to do here.

In this work, we  first perform the Schmidt basis diagonalisation for arbitrary energy eigenstates of the Hydrogenic system with a free centre-of-mass wavefunction. We find that the spread in the Schmidt basis scales with $n^{-1}$, where $n$ is the principal quantum number, which implies that  Hydrogenic systems  found in highly excited states are less entangled. These results apply to any system that can be approximated as free, which includes those confined to shallow traps with large centre-of-mass wavefunctions. To treat localised systems, we assume that the centre-of-mass wavefunction is prepared in  the shape of  a three-dimensional Gaussian wavepacket. We then take inspiration from the study of entanglement in Gaussian continuous variable states~\cite{simon2000peres, shchukin2005inseparability} to provide a sufficient entanglement test for localised systems,  which is based on the positive-partial transpose (PPT) criterion. Our results are valid for any energy eigenstates, and succeeds in detecting entanglement for a large number of parameter configurations. 

This work is structured as follows. We begin by introducing the Hydrogenic solutions to the time-independent Schr{\"o}dinger equation in Section~\ref{sec:Hydrogen:atom}. We then present the Schmidt basis diagonalisation for arbitrary energy eigenstates in Section~\ref{sec:entanglement:free:Hydrogen} for a free Hydrogenic system, and then proceed in Section~\ref{sec:localised} to define a sufficient entanglement test for a Hydrogenic system localised to a three-dimensional Gaussian wavepacket. The work is concluded by a discussion in Section~\ref{sec:discussion} and some concluding remarks in Section~\ref{sec:conclusions}.

\section{The Hydrogenic solutions} \label{sec:Hydrogen:atom}

We begin by considering two systems with a joint wavefunction $\Psi( \vec{r}_1, \vec{r}_2) $ that is parametrised by the position vectors $\vec{r}_1$ and $\vec{r}_2$. We allow the two systems to interact via a central potential of the form $V(\vec{r}) = \alpha/|\vec{r}|$, where $\vec{r}=\vec{r}_1-\vec{r}_2$ and $\alpha>0$ is a generic coupling constant that depends on the interaction\footnote{We only consider attractive potentials, which means that $V( \vec{r}_1, \vec{r}_2)$ is positive when the minus sign is explicitly included in the Schr{\"o}dinger equation in Eq.~\eqref{eq:two-particle-schordinger}}. The time-independent Schr{\"o}dinger equation that describes the system is given by 
\begin{equation} 
\left[ - \frac{\hbar^2}{2m_1} \nabla^2_1 - \frac{\hbar^2}{2m_2} \nabla^2_2 - V( \vec{r})  \right] \Psi(\vec{r}_1 , \vec{r}_2) = i \hbar \frac{\p }{\p t} \Psi( \vec{r}_1 , \vec{r}_2) .
\end{equation}
As is well known, 
this equation can be split into a relative and a centre-of-mass equation resulting from the following change of coordinates:
\begin{align}  \label{eq:definition:relative:coordinates}
\vec{R} & =  \frac{m_1 \vec{r}_1 + m_2 \vec{r}_2}{ m_1 + m_2} , \nonumber \\
\vec{r} &=   \vec{r}_1 - \vec{r}_2 , 
\end{align}
where $m_1$ and $m_2$ are the masses of the particles. 
Note that the centre-of-mass and relative momenta $\vec{P}=(\vec{p}_1+\vec{p}_2)$ and 
$\vec{p}=(m_2 \vec{p}_1- m_1 \vec{p}_2)/(m_1+m_2)$ are conjugate to $\vec{R}$ and $\vec{r}$.  The momentum coordinates must be chosen such that the canonical commutator relations are preserved at all times. See~\ref{app:Hydrogen:solutions} for details on how the separation of variables is performed. 
The separated Schr{\"o}dinger equations become
\begin{align}
 - \frac{\hbar^2}{2 M } \nabla^2_{\vec{R}} \varphi(\vec{R}) =  E_{\mathrm{CM}} \,  \varphi(\vec{R})   , \label{eq:centre:of:mass:SE} \\
\left[  -\frac{\hbar^2}{2\mu} \nabla_{\vec{r}}^2  - V(\vec{r}) \right] \psi(\vec{r}) = E_{\mathrm{rel}} \, \psi(\vec{r}) ,
\end{align}
where $\mu=m_1m_2/(m_1+m_2)$ is the reduced mass of the two-particle system, and where $M = m_1 + m_2$ is the total mass.
The two decoupled  time-independent Schr{\"o}dinger equations allow for solutions comprised of a relative wavefunction 
$\psi_{nlm}( \vec{r})$, where $n,l$, and $m$ are the principal, angular and magnetic quantum numbers,  and a centre-of-mass wavefunction $\varphi( \vec{R})$ that  takes the form of an infinite plain wave. 
Later on, we shall also consider the more realistic case  where the centre-of-mass wavepacket is localised,
and study how the entanglement is affected by its  finite volume in space.

The full state $\Psi(\vec{r}, \vec{R})$ of the system can now be written in terms of the relative coordinate $\vec{r}$ and centre-of-mass coordinate $\vec{R}$ as the following separable state:
\begin{equation} \label{eq:initial:full:state:position}
 \Psi( \vec{r}, \vec{R}) = \psi_{nlm} ( \vec{r} ) \, \varphi( \vec{R}) .
\end{equation}
The Hydrogenic wavefunction $\psi_{nlm}(\vec{r})$  is given by 
\begin{equation} \label{eq:initial:state:Hydrogen:position}
\psi_{nlm}( \vec{r}) =  R_{nl}(r) \, Y_l^m( \theta, \phi), 
\end{equation}
where  $R_{nl}(r)$ is the radial wavefunction, which depends on the Laguerre polynomials, and $Y_l^m(\theta, \phi)$ are the spherical harmonics. These functions are described in full in~\ref{app:Hydrogen:solutions}. 

While the state in Eq.~\eqref{eq:initial:full:state:position} is separable in the $\{\vec{r}, \vec{R}\}$ basis, it cannot be written as a separable state in the original $\{\vec{r}_1, \vec{r}_2\}$ basis.  As a result, the two subsystems are entangled, and it remains to determine how such entanglement can be detected. 

Before we proceed to examine the entanglement of the system, we present the momentum representation of the Hydrogenic wavefunctions, since these will be of great use to us later. They correspond to the Fourier transform of $\psi_{nlm}( \vec{r})$, and are given by~\cite{bransden2003physics}:
\begin{equation} \label{eq:Hydrogen:general:wavefunction:momentum}
\tilde{\psi}_{nlm}(\vec{k})  = F_{nl}(k) \, Y_l^m (\theta, \phi), 
\end{equation}
where $F_{nl}(k)$ is a function of the Gegenbauer polynomials~\cite{erdelyi1955higher}, and $Y_l^m (\theta,\phi)$ are again the spherical harmonics,  and $(k, \theta, \phi)$ are the wavevector coordinates.  More details on this solution and especially the Gegenbauer polynomials can again be found in~\ref{app:Hydrogen:solutions}.

\section{Entanglement of the free Hydrogen atom} \label{sec:entanglement:free:Hydrogen}
We start by considering the case when the centre-of-mass wavefunction $\varphi(\vec{R})$ in Eq.~\eqref{eq:initial:full:state:position} corresponds to that of a free wavefunction (see Figure~\ref{fig:free}):
\begin{equation}
\varphi( \vec{R}) = \frac{1}{\sqrt{V}} e^{- i \vec{K} \cdot \vec{R}}, 
\end{equation}
where $V$ is the free space normalisation volume, which we take to infinity at the end of each calculation, $\vec{R}$ is given in Eq.~\eqref{eq:definition:relative:coordinates} and $\vec{K} = (\vec{k}_1 + \vec{k}_2)$ is the centre-of-mass wavevector of the state. The full eigenstate, including the Hydrogenic part is then given by
\begin{equation} 
\Psi( \vec{r}_1, \vec{r}_2) = \Psi(\vec{r}, \vec{R}) = \psi_{nlm}(\vec{r})  \frac{1}{\sqrt{V} }e^{-i \vec{K} \cdot \vec{R}}.
\end{equation}
For all $\vec{K}$, one can perform a Galillean transformation into a frame where the centre-of-mass is stationary, 
corresponding to the product of local unitary operations 
${\mathrm e}^{i m_1\vec{K}\cdot\vec{r}_1/(m_1+m_2)} \otimes {\mathrm e}^{i m_2\vec{K}\cdot\vec{r}_2/(m_1+m_2)}$. 
As one should expect, the entanglement cannot therefore depend on $\vec{K}$, which can be set to zero without 
loss of generality.

Transforming into a frame with $\vec{K} = 0$  leaves us with the following state:
\begin{equation} \label{eq:Hydrogen:com:frame}
\Psi( \vec{r}, \vec{R}) = \frac{1}{\sqrt{V}} \psi_{nlm}( \vec{r}) .
\end{equation}
In the remainder of this section, we compute the Schmidt basis associated with these eigenstates and generalise the 
results of Ref.~\cite{tommasini1998hydrogen} to arbitrary energy eigenstates. We begin by setting out a few preliminaries 
concerning the Schmidt basis decomposition for continuous variable systems.

\subsection{The Schmidt basis} \label{sec:Schmidt:basis}
A necessary and sufficient test to verify whether a pure state is entangled or not consists in casting it in its Schmidt basis 
and  computing its Schmidt rank.  If the Schmidt rank is greater than one, the state 
is entangled. 
The Schmidt basis and the Schmidt rank are  straight-forward to introduce for discrete systems, but they have also been studied for continuous variable systems in full generality~\cite{parker2000entanglement}. We here provide a short review of the Schmidt diagonalisation for discrete systems and  its generalisation to continuous systems,  with particular  
reference to the wavefunction notation we are adopting.

It may be shown that a choice of local bases exist such that an arbitrary bipartite state $\ket{\Psi}$ can be expanded  
in a basis formed by pairs of distinct, orthonormal local vectors, which we shall call $\ket{w_k}=\ket{u_k}\otimes\ket{v_k}$: 
\begin{equation} \label{eq:diagonalised:Schmidt:state:discrete}
\ket{\Psi} = \sum_k c_k \ket{w_k} = \sum_k c_k (\ket{u_k}\otimes\ket{v_k}),
\end{equation}
where the Schmidt coefficients $c_k$ may be taken  to be positive and real. The local Schmidt bases are nothing but the 
eigenbases of the local density operators.

It is interesting to note that the local state of a subsystem, defined as $\hat \varrho_1 = \mathrm{Tr}_2 [ \ket{\Psi}\bra{\Psi}]
= \sum_{k}c_k^2 \ket{u_{k}}\bra{u_k}$, satisfies the following eigenvalue equation:
\begin{equation} \label{eq:discrete:Schmidt:basis:eigenvalue:equation}
(\hat \varrho_1\otimes \hat{\mathbbm{1}}_2) \ket{w_k} = {c}_k^2 \ket{w_k}. 
\end{equation}
Sometimes it is easier to solve this equation for the basis $\ket{w_k}$, and thereby reconstruct the state in 
Eq.~\eqref{eq:diagonalised:Schmidt:state:discrete} than to perform the usual Gram--Schmidt diagonalisation procedure. 

The link between  wavefunction notation and bra-ket notation is the inner product, where the wavefunctions are scalar functions of the state space, allowing to represent the latter in the (improper) eigenvectors of the position operator. For example:
\begin{equation}
\psi(x) = \braket{x |\psi }, 
\end{equation}
where $\ket{x}$ are the position eigenstates and $\ket{\psi}$ is the state of the system. 

In order to introduce the continuous analogue of the Schmidt basis note that, given the wavefunction for a bipartite state $\Psi(x,y)$, where $x$ and $y$ denote the positions of the subsystems (assumed here to be one-dimensional, for simplicity), the analogue of the density matrix is given by 
\begin{equation}
\varrho(x', y', x, y) = \Psi^*(x', y') \Psi(x, y), 
\end{equation}
which is normalised as follows: 
\begin{equation}
\int \mathrm{d}x \int \mathrm{d}y \, \varrho(x, y, x, y) = 1. 
\end{equation}
The density matrix in the position representation $\varrho_1(x', x)$ of
the traced-out subsystem is then given by 
\begin{equation}
\varrho_1(x', x) = \int \mathrm{d}y \, \Psi^*(x',y) \Psi(x,y).
\end{equation}
We concluded before that finding the Schmidt basis for discrete states amounts to solving the eigenvalue equation in Eq.~\eqref{eq:discrete:Schmidt:basis:eigenvalue:equation}. The continuous analogue of the eigenvalue problem is the following integral equation~\cite{parker2000entanglement}, otherwise known as a Fredholm equation of the second kind~\cite{arfken1999mathematical}:
\begin{equation} \label{eq:Fredholm:equation}
\int\mathrm{d}x' \,   \varrho_1 ( x',x) \, \phi_i (x')  = \lambda_i  \, \phi _i(x) \, , 
\end{equation}
where $\{\lambda_i \}$ is the set of eigenvalues of  $\varrho_1(x', x)$ parametrised by the index $i$. 

Since we in this work are interested in bipartite systems in three spatial dimensions, we write the density matrix of the free Hydrogenic state in Eq.~\eqref{eq:Hydrogen:com:frame} as
\begin{equation}
\varrho(\vec{r}_1', \vec{r}_2', \vec{r}_1, \vec{r}_2) = \Psi^*(\vec{r}_1', \vec{r}_2') \,  \Psi(\vec{r}_1, \vec{r}_2). 
\end{equation}
This allows us to rewrite the Fredholm equation in Eq.~\eqref{eq:Fredholm:equation} in terms of a three-dimensional subsystem state $\varrho_1(\vec{r}_1', \vec{r}_1)$, given by 
\begin{equation}\label{defloco}
\varrho_1( \vec{r}_1', \vec{r}_1) = \int \mathrm{d}\vec{r}_2 \, \Psi^*( \vec{r}_1', \vec{r}_2) \,  \Psi(\vec{r}_1, \vec{r}_2), 
\end{equation}
whence  the Fredholm equation becomes
\begin{equation}\label{eq:continuous:Fredholm:eq:3D}
\int \mathrm{d} \vec{r}_1' \, \varrho_1( \vec{r}_1', \vec{r}_1) \, \phi_{\vec{k}} (\vec{r}_1') = \lambda_{\vec{k}} \, \phi_{\vec{k}} ( \vec{r}_1), 
\end{equation}
where we have indexed $\phi_{\vec{k}}$ with the (possibly) continuous index $\vec{k}$. 
In terms of the full state, the Fredholm equation in Eq.~\eqref{eq:Fredholm:equation} becomes
\begin{equation} \label{eq:continuous:Fredholm:eq:3D:state}
\int  \mathrm{d}\vec{r}_1'  \,   \mathrm{d}\vec{r}_2 \, \Psi^*( \vec{r}_1', \vec{r}_2) \,  \Psi( \vec{r}_1, \vec{r}_2) \, \phi_{\vec{k}} (\vec{r}_1')= \lambda_{\vec{k}} \, \phi _{\vec{k}}(\vec{r}_1) \, . 
\end{equation}
To obtain the diagonalisation in three dimensions, Eq.~\eqref{eq:continuous:Fredholm:eq:3D:state} must be solved for $\phi_{\vec{k}}$. We anticipate that finding the solution in general will be an extremely challenging task. In the next Section, however, we show that Eq.~\eqref{eq:continuous:Fredholm:eq:3D:state} has a simple solution when the system is invariant under spatial translations.

\subsection{Schmidt basis diagonalisation of a  translationally invariant state} \label{sec:translation:invariant:diagonalisation}
Before considering the Schmidt diagonalisation of the free Hydrogenic wavefunctions, we proceed with a proof showing that  \textit{translationally invariant} (``\textit{homogeneous}'') states are diagonalised by the Fourier transform, which was first demonstrated in Ref.~\cite{tommasini1998hydrogen}. We define translationally invariant by the fact that displacing the two subsystems by the same distance in space leaves the wavefunction invariant, up to a global phase. 

A translationally invariant wavefunction must depend on the variable $\vec{r}$ alone, and not on $\vec{R}$, 
since the former is invariant under an equal translation of $\vec{r}_1$ and $\vec{r}_2$ whilst the latter may be modified arbitrarily by such translations. 
This implies that the local density function ${\varrho}_1(\vec{r}_1',\vec{r}_1)$, whose eigenbasis determines the Schmidt decomposition, is a function only of the difference of its arguments, since Eq.~\eqref{defloco} in this case reads 
\begin{align}
\varrho_1( \vec{r}_1', \vec{r}_1) &= \int \mathrm{d}\vec{r}_2 \, \Psi^*( \vec{r}_1'-\vec{r}_2) \Psi(\vec{r}_1-\vec{r}_2)  \\
&= \int \mathrm{d}\vec{r}_2 \, \Psi^*(\vec{r}_2) \Psi( \vec{r}_1-\vec{r}_1'+\vec{r}_2) = \varrho_1( \vec{r}_1' - \vec{r}_1) , \nonumber
\end{align}
where we have made the variable substitution $\vec{r}_2 \rightarrow \vec{r}_1'- \vec{r}_2$, treating $\vec{r}_1'$ as a constant. 
The Fourier transform and inverse of $\varrho_{1}(\vec{r}_1-\vec{r}_1')$ may then be expressed in terms of a single 
(three-dimensional) variable:
\begin{align}
\tilde{\varrho}(\overbar{\vec{k}}) &=  \int \mathrm{d} \overbar{\vec{r}} \,  e^{- i \overbar{\vec{k}} \cdot \overbar{\vec{r}} }\,\varrho( \overbar{\vec{r}}) ,  \label{eq:fourier:transform}\\
\varrho( \overbar{\vec{r}}) &= \frac{1}{( 2\pi)^{3}} \int \mathrm{d} \overbar{\vec{k}} \, e^{i \overbar{\vec{k}} \cdot \overbar{\vec{r}}} \, \tilde{\varrho}(\overbar{\vec{k}}),  \label{eq:fourier:transform:inverse}
\end{align} 
where we have defined $\overbar{\vec{r}} = (\vec{r}_1 - \vec{r}_1')$ and its conjugate variable  $\overbar{\vec{k}} = (\vec{k}_1 - \vec{k}_1')$ \footnote{Note that we have added the bar to the local variables to differentiate them from the relative and centre-of-mass position and momentum variables introduced earlier}.

We then recall the Fredholm eigenvalue equation for continuous systems given in Eq.~\eqref{eq:continuous:Fredholm:eq:3D}. Inserting the translationally invariant state and the momentum eigenstate ansatz 
$\phi_{\overbar{\vec{k}}} = e^{i \overbar{\vec{k}}\cdot \vec{r}_1'}$, we find:
\begin{align}
\int \mathrm{d}\vec{r}_1' \,   \varrho_1 ( \vec{r}_1 - \vec{r}_1')  \, e^{i  \, \overbar{\vec{k}} \cdot \vec{r}_1'} &= \frac{1}{( 2\pi)^{3}} \int \mathrm{d} \overbar{\vec{k}}' \,  \tilde{\varrho}( \overbar{\vec{k}}') \, e^{i \, \overbar{\vec{k}}' \cdot \vec{r}_1 } \int \mathrm{d} \vec{r}_1' \, e^{i \left( \overbar{\vec{k}} - \overbar{\vec{k}}'\right) \cdot \vec{r}_1'} = \tilde{\varrho}( \overbar{\vec{k}}) \, e^{i \, \overbar{\vec{k}} \cdot \vec{r}_1},
 \end{align}
where we have used the Fourier transform of $\varrho(\overbar{\vec{r}})$ shown in Eq.~\eqref{eq:fourier:transform:inverse}.
This demonstrates that the local density operator is diagonal in the local momentum eigenbasis, which thus form its 
associated Schmidt basis, with Schmidt coefficients given by the Fourier transform 
$\tilde{\varrho}(\overbar{\vec{k}})$.  In other words, a translationally invariant state is always diagonalised by its Fourier transform.

\subsection{Entanglement for general Hydrogenic energy eigenstates}  \label{sec:free:Hydrogen:general:eigenstates}
We have shown that the local density operator of homogeneous state is always diagonalised by the Fourier transform. A free Hydrogenic system, whose wavefunction is shown in Eq.~\eqref{eq:Hydrogen:com:frame}, fulfills the criterion of translation invariance since it depends solely on the distance $\vec{r} = (\vec{r}_1  -\vec{r}_2)$ between the two systems. 

In order to compute the Schmidt coefficients of a Hydrogenic eigenfunction, and hence qualify its entanglement,
we need first to trace out one of the two particles, as follows 
\begin{align}
\varrho_1( \vec{r}_1-\vec{r}_1') &= \int \mathrm{d} \vec{r}_2 \, \Psi^*( \vec{r}_1', \vec{r}_2) \Psi( \vec{r}_1, \vec{r}_2) \nonumber \\
&= \frac{1}{V} \int \mathrm{d} \vec{r}_2 \, \psi_{nlm}^*( \vec{r}_1' - \vec{r}_2)  \psi_{nlm}( \vec{r}_1 - \vec{r}_2) \nonumber \\
&=  \frac{1}{V}\int \mathrm{d} \vec{y} \, \psi_{nlm}^*(  \vec{y}) \psi_{nlm}(\overbar{\vec{r}}+\vec{y}). 
\end{align} 
Here, through the substitution $\vec{y} = \vec{r}_1' - \vec{r}_2$, we have again shown that the traced-out Hydrogenic state is translation invariant, since it only depends on the local variable $\overbar{\vec{r}} = \vec{r}_1 - \vec{r}_1'$. As we proved in the previous Section, the Fourier transform $\tilde{\varrho}_1(\overbar{\vec{k}})$ corresponds to the Schmidt basis of the Hydrogenic system, where $\overbar{\vec{k}} = \vec{k}_1 - \vec{k}_1'$ is the Fourier complement to $\overbar{\vec{r}}$.

The state is separable if and only if the probability distribution $\tilde{\varrho}(\overbar{\vec{k}})$ is a delta-function, which 
provides one with a criterion to test entanglement in this context. This corresponds to the discrete analogue of having one single non-zero Schmidt coefficient.

Notice that, even if the continuous analogue of the Schmidt coefficients are known, 
an actual quantification of the entanglement is 
difficult in this case.  The local von Neumann entropy is in fact not a good quantifier for such continuous systems as, being the Shannon entropy of the local state's 
spectrum, it is ill defined if the latter is continuous, as is always the case for our homogeneous systems, which are diagonal in the (continuous) local momentum basis. The differential entropy \cite{cover2012elements}, sometimes introduced as the continuous analogue of the von Neumann entropy, is problematic (it can even be negative!), and lacks the properties to qualify as a bona fide entanglement monotone\footnote{The differential entropy should not be confused with the \emph{relative entropy}, for which the discrete and continuous version have identical properties}. It may even be shown that, in infinite dimension, states with diverging von Neumann entropy may be found arbitrarily close, in the trace norm topology, to \emph{any} quantum state~\cite{eisert2002quantification}. 

The linear entropy $S_{\rm{Lin}}$ is sometimes favoured as an entanglement quantifier because it is easier to compute. However, besides being an unwieldy quantity in our case, the local linear entropy is not endowed with an operational interpretation. 

It will therefore be convenient to adopt a different quantifier of entanglement, which allows us to illustrate the behaviour of Hydrogenic entanglement more clearly. We opt to evaluate an entanglement quantifier already suggested 
in~\cite{tommasini1998hydrogen}, namely one given by the standard deviation of the Schmidt function:
\begin{equation} \label{eq:standard:dev:Schmidt:basis}
\Delta \overbar{k} = \sqrt{\braket{\overbar{k}^2} - \braket{\overbar{\vec{k}}}^2}.
\end{equation} 
Eq.~\eqref{eq:standard:dev:Schmidt:basis} does in a sense quantify the deviation from the delta function that characterises separable states:
if $\Delta \overbar{k} = 0$, the state is separable, and if $\Delta \overbar{k} > 0$, the state is entangled. 

We proceed to compute $\Delta \overbar{k}$. The continuous analogue of computing expectation values with the trace operation is given by 
\begin{align}
\braket{\bar{\vec{k}}} 
&=\int \mathrm{d} \vec{r}_1 \int \mathrm{d} \vec{r}_1' \, \delta( \vec{r}_1- \vec{r}_1') \left( - i \boldsymbol{\nabla}_{(\vec{r}_1 - \vec{r}_1')} \right) \varrho_1( \vec{r}_1 - \vec{r}_1').
\end{align}
We then insert the Fourier transform in Eq.~\eqref{eq:fourier:transform:inverse} to find
\begin{align}
\braket{\bar{\vec{k}}} &= - \frac{i}{(2\pi)^3} \int \mathrm{d} \vec{r}_1 \int \mathrm{d} \vec{r}_1' \,\int \mathrm{d} \overbar{\vec{k}}   \, \delta( \vec{r}_1 - \vec{r}_1') \,  \tilde{\varrho}_1 ( \overbar{\vec{k}}) \boldsymbol{\nabla}_{(\vec{r}_1 - \vec{r}_1')}   e^{i \bar{\vec{k}} \cdot \left( \vec{r}_1 - \vec{r}_1'\right)}  \nonumber \\
&= \frac{V}{(2\pi)^{3}} \int \mathrm{d} \overbar{\vec{k}} \, \overbar{\vec{k}} \,  \tilde{\varrho}_1(\overbar{\vec{k}}), 
\end{align}
where  the integration volume $V$ appears because we integrate over all space. 
This expectation value $\braket{\bar{\vec{k}}}$ is calculated by integrating over all vectors, which averages to zero: 
\begin{align}
\braket{\overbar{\vec{k}}} &= \frac{V}{(2\pi)^{3}}\int \mathrm{d}\overbar{\vec{k}} \,  \overbar{\vec{k}} \,  \tilde{\varrho}_1(\overbar{\vec{k}}) =0. 
\end{align}
The variance $\braket{\overbar{k}^2}$, on the other hand, is given by 
\begin{align}
\braket{\overbar{k}^2} &=\frac{V}{(2\pi)^{3}} \int \mathrm{d} \overbar{\vec{k}} \, \overbar{k}^2 \, \tilde{\varrho}_1( \overbar{\vec{k}}) . 
\end{align}
Using the momentum representation of the Hydrogenic wavefunctions,  shown in Eq.~\eqref{eq:Hydrogen:general:wavefunction:momentum},   and realising that the free system is proportional to $V^{-1}$, we find
\begin{align}
\braket{\overbar{k}^2} &= \frac{1}{(2\pi)^3} \int^\infty_0 \mathrm{d}\overbar{k}  \, \overbar{k}^4 \,   [F_{nl}(\overbar{k})]^2   \int^{\pi}_0 \mathrm{d}\theta \sin{\theta} \, \int^{2\pi}_0 \mathrm{d}\phi \,   |Y^m_l(\theta, \phi) |^2. 
\end{align}
The integrals over $\theta$ and $\phi$ satisfy the normalisation condition for the spherical harmonics (see Eq.~\eqref{app:eq:spherical:harmonics:normalisation} in~\ref{app:Hydrogen:solutions}), and we are left with 
\begin{align}
\braket{\bar{k}^2}
&= \frac{1}{(2\pi)^3} \int_0^\infty \mathrm{d}\bar{k} \,\bar{k}^4  \, |F_{nl}(\bar{k})|^2.
\end{align}
The integral over $\bar{k}$ returns a well-known  result from atomic physics. It admits the standard solution~\cite{bransden2003physics}:
\begin{align} \label{eq:momentum:variance:result}
\braket{\bar{k}^2}&=  \frac{1}{(2\pi)^3}  \frac{1}{n^2a_0^2} \, .
\end{align}
We prove this relation, which extends what was previously known about entanglement for the Hydrogenic ground state to any energy eigenstate
\cite{tommasini1998hydrogen}, in~\ref{app:sec:Hydrogen:wavefunction:momentum:variance}.

We conclude that the standard deviation of the local momentum, 
related to the local wavevector by $\bar{p} = \hbar \bar{k} $, reads 
\begin{align} \label{eq:Schmidt:rank:free:Hydrogen}
\Delta \bar{p}  =  \frac{1}{(2\pi)^{3/2}}\frac{\hbar}{n a_0} = \frac{1}{(2\pi)^{3/2}} \frac{\alpha \mu}{n \hbar},
\end{align}
where the expression for the reduced Bohr radius $a_0=\hbar^2/(\alpha \mu)$ was inserted 
(recall  that $\mu$ is the reduced mass of the system and that $\alpha$  is the potential interaction strength, which has dimensions of an energy times a length). 

The standard deviation of the continuous local spectrum only depends on the principal quantum number $n$, and not 
on $l$ and $m$. This is the case since the operations that transform sectors with the same $n$ and different $l$ and $m$ 
are local unitary operations (which is particularly clear for the azimuthal number $m$, whose different values are 
related by rotations), and thus cannot affect properties related to the Schmidt spectrum and entanglement. 

On the other hand, the distribution of local momentum is more spread out for lower principal quantum numbers, which 
is non-trivial and suggests that,
in practice, lower quantum states would be more favourable for the detection of quantum correlations in  Hydrogenic systems.  
Notice that the fact that we can qualify 
the presence or absence of entanglement so liberally (by any positive value of the standard deviation!) 
is just an artefact of the fact that our theoretical finding applies to ideal (pure) eigenstates.

Further, our quantifier grows with the interaction strength, as one should expect, and with the reduced mass, 
which indicates that, {\em at a specific given total mass}, the entanglement is maximum for two equally distributed masses. However, the eigenstate of a Hydrogen atom would be about twice more entangled than the corresponding 
positronium eigenstate, since the large mass of the proton nucleus factors out of $\mu$ and leaves $\mu = m_{\mathrm{e}}$, where $m_{\mathrm{e}}$ is the electron mass. Compare this with the reduced mass of the positronium eigenstate (a bound state of one electron and one positron), which reads $\mu = m_{\mathrm{e}} m_{\mathrm{p}}/(m_{\mathrm{e}} + m_{\mathrm{p}}) = m_{\mathrm{e}}/2$. We also note that the expression in Eq.~\eqref{eq:Schmidt:rank:free:Hydrogen} is not independent of the choice of Fourier normalisation, however, as this is merely a scaling factor, the choice does not significantly affect the results.

Before we proceed to investigate localised Hydrogenic states, let us briefly return to the linear entropy $S_{\mathrm{Lin}}$ of a Hydrogenic system. We find that the linear entropy converges for a free Hydrogenic systems, although, as mentioned above, it does not admit a direct operational interpretation as a quantifier of the entanglement. We merely provide this result for completeness. 

In~\ref{app:linear:entropy}, we show that the linear entropy $S_{\mathrm{Lin}}$ for a free Hydrogenic system is given by  
\begin{equation} \label{eq:compact:linear:entropy}
S_{\mathrm{Lin}} = 1 -  \frac{1}{V} \int \mathrm{d} \bar{\vec{k}} \left[ F_{nl} (\bar{k}) \right]^4 \left[ Y_{lm}(\theta, \phi) \right]^4 . 
\end{equation}
While we find that the integral in Eq.~\eqref{eq:compact:linear:entropy} converges (see the rather lengthy expression in Eq.~\eqref{app:eq:final:expression:linear:entropy}), we conclude that when $V\rightarrow \infty$, the linear entropy tends to $S_{\mathrm{Lin}} = 1$ for all energy eigenstates.

\section{Entanglement of a localised Hydrogenic system} \label{sec:localised}
 The approximation of the system as free breaks down when the overall centre-of-mass wavepacket is of a similar width compared with the characteristic length scale  $a_0$ of the system. For this setting, the methods and results demonstrated in the previous section are no longer valid. By definition, a localised state is no longer invariant under spatial translations, which means that it can no longer be diagonalised by the Fourier transform. Finding the Schmidt basis of the state would require solving the integral equation shown in Eq.~\eqref{eq:continuous:Fredholm:eq:3D:state}, which we anticipate to be an extremely difficult task. 

In this work, we choose instead to explore an entanglement test
based on detecting a violation of the positivity of the partial transpose through the 
information contained in the second statistical moments of the localised Hydrogenic state~\cite{simon2000peres}.
 One should bear in mind that such a test applied to non-Gaussian states
is only sufficient, and not necessary for entanglement.

We consider therefore the case where the centre-of-mass wavepacket  is a three-dimensional  Gaussian wavepacket centred at the origin, as shown in Figure~\ref{fig:Gaussian}. While a Gaussian wavepacket is not a solution to the time-independent 
Schr{\"o}dinger equation of a free particle, it is nonetheless a very reasonable localised state, which can be prepared 
and maintained, for instance, by trapping the system in a harmonic potential that addresses the centre-of-mass degree of freedom. The addition of dynamics into this setting is a problem we leave for future work.

The Gaussian wavefunction  of the centre-of-mass position variable $\vec{R}$ is given by 
\begin{equation} \label{eq:Gaussian:wavefunction}
\varphi( \vec{R} ) = \frac{1}{\pi^{3/4} \, b^{3/2}} e^{- \vec{R} \cdot \vec{R}/(2b^2)}, 
\end{equation}
where $b$ has units of length and encodes the standard deviation of the wavepacket at the specific moment in time for which we consider the system. The full state of a localised Hydrogenic system therefore becomes
\begin{equation} \label{eq:localised:Gaussian:Hydrogen}
\Psi( \vec{r}, \vec{R}) = \psi( \vec{r}) \, \varphi( \vec{R})  .
\end{equation}
It is clear that the state appears entangled even in the Gaussian wavepacket alone, since $\varphi(\vec{R})$ in 
Eq.~\eqref{eq:Gaussian:wavefunction} cannot be written as a product of functions of $\vec{r}_1$ and $\vec{r}_2$. We therefore expect contributions to entanglement from both the relative and centre-of-mass degrees of freedom.

We proceed now to study the entanglement criterion for this state. For simplicity, we restrict the analysis to the case of equal masses, such that $m_1 = m_2$. 

\subsection{Entanglement in the first and second moments} \label{sec:entanglement:second:moments}
In order to devise an entanglement test for a localised Hydrogenic state, we adopt  some well-established techniques 
developed in the study of quantum information with continuous variables~\cite{serafini2017quantum}. 
In particular, a necessary and sufficient entanglement 
criterion based on the second order statistical moments of the canonical operator exist 
for two-mode Gaussian states~\cite{simon2000peres,duan2000inseparability, lami2018gaussian}, which is however sufficient 
to detect the entanglement of any quantum state\footnote{Indeed, a non-Gaussian state that is entangled might not appear entangled through examination of its second moments alone. As already mentioned, the test we will adopt is therefore \textit{sufficient} but not \textit{necessary} for determining entanglement.}.

Such a criterion is based on a general necessary test for the positivity of the partial transpose (PPT) at the level of second moments, and on the observation that, in turn, PPT is necessary for the separability of arbitrary states~\cite{peres1996separability,horodecki1997separability}. A necessary criterion 
for separability is equivalent to a sufficient criterion for its converse, namely quantum entanglement. 

An entanglement test on the second moments can be carried out as follows. Given a suitable canonical 
basis of self-adjoint operators $\hat{\vec{r}} = ( \hat x_1, \hat p_1, \hat x_2, \hat p_2 \ldots \hat x_ N , \hat p_N)^{\mathrm{T}}$ for $ N$ modes, 
the covariance matrix $\sigma$ is given by 
\begin{equation}
\sigma = \mathrm{Tr} \left[  \{\hat{\vec{r}}, \hat{\vec{r}}^{\mathrm{T}} \} \, \hat \varrho \right] , 
\end{equation}  
where $\hat \varrho$ is a Gaussian (or non-Gaussian) state, and where the transpose $\mathrm{T}$ is taken in the outer-product sense, including all possible pairs of operators to form a real, symmetric matrix of correlations. 

For continuous variable systems, partial transposition is equivalent to changing the sign of the second canonical variable. This implies that half of the off-diagonal elements in the covariance matrix, namely all elements that contain an odd power of the second momentum variable, gain a minus sign~\cite{serafini2017quantum} (we demonstrate this explicitly in~\ref{app:sec:covariance:matrix}). 
Any physical (i.e., derived from a trace-class, positive density operator) covariance matrix $\sigma$
must satisfy the uncertainty principle that can be cast as $-(\Omega \,  \sigma^{\mathrm{Tp}})^2\ge \hat{\mathbbm{1}}$, 
where $\Omega$ is the symplectic form defined in this basis as
\begin{align} \label{eq:symplectic:form:definition}
\Omega =\bigoplus_{j=1}^n \,  \omega \, ,  \,  \mbox{ with  } \quad &\omega =   \left( \begin{matrix} 0 & 1 \\
- 1 & 0 \end{matrix} \right) \, .
\end{align}
The sufficient entanglement test can thus be stated as
\begin{equation}
\exists \; j\; : \;\tilde{\nu}_j < 1 \; ,
\end{equation}
where $\tilde{\nu}_j$ are the $N$ symplectic eigenvalues of the partially transposed covariance matrix $\sigma^{\mathrm{Tp}}$. The symplectic eigenvalues are defined as the square roots of the eigenvalues of  
$- (\Omega \, \sigma^{\mathrm{Tp}})^2$, which are at least two-fold degenerate (so that there are $N$ independent ones
in the case on hand). 

Hence, in order to demonstrate entanglement for the Hydrogenic  wavefunction (which are globally non-Gaussian due to the relative coordinate contribution, even if the centre-of-mass coordinate is taken in a Gaussian wave-packet, which in turn would amount to a Gaussian state in the quantum optics nomenclature), it suffices to show that one of the symplectic eigenvalues of its covariance matrix satisfies $\tilde{\nu}_j<1$. We therefore need to evaluate its second moments, which is what we do in the following section.

\subsection{Second moment entanglement for Hydrogenic states}

The bipartite state in Eq.~\eqref{eq:localised:Gaussian:Hydrogen} has support in three spatial dimensions. This setting yields six independent modes (one for each spatial direction for each of the subsystems). It follows that the covariance matrix is a $12\times 12 $ matrix (with two variables per mode). To construct the matrix, we must compute the expectation values and variances for the position and momentum variables in each mode.

In the original partition in terms of the $\vec{r}_1$ and $\vec{r}_2$ coordinates  for the original two subsystems, we construct the following (dimensionfull) basis of coordinates:
\begin{equation} \label{eq:Gaussian:original:basis}
\vec{\mathbb{X}} = \left( x_1,\, p_{x1}, \, y_1, \, p_{y1},  \, z_1, \, p_{z1}, \, x_2, \, p_{x 2}, \, y_2, \, p_{y 2}, \, z_2, \, p_{z 2} \right)^{\mathrm{T}}.
\end{equation}
It would be cumbersome to compute the first and second moments directly in the basis $\mathbb{X}$ of Eq.~\eqref{eq:Gaussian:original:basis}, since the variables $\vec{r}_1$ and $\vec{r}_2$ are mixed between the Hydrogenic eigenstates and the Gaussian wavepacket. 
However, we can simply evaluate the statistical moments in the $\{\vec{r}, \vec{p},\vec{R}, \vec{P}\}$ basis, 
separately for the Hydrogenic wavefunction and the Gaussian wavepacket, 
and then apply the symplectic transformation $S$ that  relates the local coordinates to the decoupled  ones on the 
covariance matrix itself, which transforms, by congruence, as $\sigma' = S \sigma S^{\mathrm{T}}$. 
With the  $S$ shown in Eq.\eqref{app:eq:symplectic:transformation} in~\ref{app:sec:covariance:matrix}, the transformed basis vector in Eq.~\eqref{eq:Gaussian:original:basis} becomes
\begin{equation}
\mathbb{X}' = \left(x, \, p_x, \,  y, \,  p_y, \,  z,  \, p_z,  \, X, \,  P_X,  \, Y,  \, P_Y,  \, Z, \,  P_Z \right)^{\mathrm{T}}\, ,
\end{equation}
where $x,y,z$ and $p_x, p_y, p_z$ are the relative position and momentum coordinates given by (for equal masses)
\begin{align}
&x = {x_1 - x_2}, && \mbox{and} && p_x = \frac{p_{x1}  - p_{x2}}{2} , 
\end{align}
and so on. Similarly, $X,Y,Z$ and $P_X, P_Y, P_Z$ are the centre-of-mass position and momentum coordinates given by 
\begin{align}
&X = \frac{x_1 + x_2}{2}, &&\mbox{and} && P_X = {p_{x1} + p_{x2}}, 
\end{align}
and so on. It is possible to find a symplectic transformation that takes unequal masses into account, but we do not do so here.

We are now ready to compute the first and second moments of the Hydrogenic states. To ensure that they are dimensionless, we rescale the position coordinates by the characteristic length scale, the reduced Bohr radius $a_0$ (see~\ref{app:Hydrogen:solutions}), and the momentum coordinates by the characteristic momentum $\hbar/a_0$. 

Using identities that involve  the Legendre polynomials and known relations such as the Kramer--Pasternak relation~\cite{pasternack1937mean,kramers1939quantentheorie}, we obtain the following first and second moments in the $\{\vec{r}, \vec{R}\}$ basis.  See~\ref{app:localised:Hydrogen} for the full, rather lengthy calculations. The  dimensionless  (where we have chosen to normalise all quantities with respect to the reduced Bohr radius $a_0$) expectation values and variances for the relative position variables of the Hydrogenic eigenstates are
\begin{align}
&\braket{x/a_0} = \braket{y/a_0} = \braket{z/a_0} = 0, 
\end{align}
and, since $\braket{x^2/a_0^2} = \braket{y^2/a_0^2}$:
\begin{align} \label{eq:Hydrogen:position:variances}
\braket{x^2/a_0^2} &=  \frac{n^2 (5n^2 - 3l(l+1) + 1)}{2} \frac{l^2+l+m^2-1}{ (2 l-1) (2 l+3)},  \\
\braket{z^2/a_0^2} &=  \frac{n^2 (5n^2 - 3l(l+1) + 1)}{2}\frac{1-2 l^2-2 l+2 m^2}{3 -4 l^2- 4 l}\nonumber.
\end{align}
The expectation values of the relative momentum variables are
\begin{align}
&\braket{p_x \,a_0/\hbar} = \braket{p_y \, a_0/\hbar} = \braket{p_z \, a_0/\hbar } = 0 , 
\end{align}
and the variances are given by, with $\braket{p_x^2\, a_0^2/\hbar^2} = \braket{p_y^2 \,a_0^2/\hbar^2}$:
\begin{align} \label{eq:Hydrogen:momentum:variances}
&\braket{p_x^2 \,a_0^2/\hbar^2}  = \frac{1}{ n^2}   \frac{l^2+l+m^2-1}{  (2 l-1) (2 l+3)}, \nonumber \\
&\braket{p_z^2\,a_0^2/\hbar^2} =  \frac{1}{n^2} \frac{1-2 l^2-2 l+2 m^2}{ 3 - 4 l^2-4 l}. 
\end{align}
As for the Gaussian wavepacket, the centre-of-mass  expectation values and variances 
read
\begin{align}
&\braket{X/a_0} = \braket{Y/a_0} = \braket{Z/a_0} = 0 \, ,\nonumber \\
&\braket{X^2/a_0^2} = \braket{Y^2/a_0^2} = \braket{Z^2/a_0^2} = \frac{b^2}{2  a_0^2}\, , \nonumber \\
&\braket{P_X \,a_0/\hbar} = \braket{P_Y\,a_0/\hbar} = \braket{P_Z\,a_0/\hbar} = 0\, ,\nonumber \\
& \braket{P_X^2\,a_0^2/\hbar^2} = \braket{P_Y^2\,a_0^2/\hbar^2} = \braket{P_Z^2\,a_0^2/\hbar^2} =\frac{a_0^2}{2 b^2} \, .
\end{align}
We now promptly transform back to the original basis $\mathbb{X} = S^{\mathrm{T}} \mathbb{X}'$, and  $\sigma = S^{-1} \sigma' S^{-1\,\mathrm{T}}$. We then apply the PPT criterion and compute the symplectic eigenvalues $\tilde{\nu}_j$ of $\sigma^{\mathrm{Tp}}$. They are given by, in terms of the previously computed variances, 
\begin{align} \label{eq:symplectic:eigenvalues}
\tilde{\nu}_1 &=    \sqrt{   \braket{x^2/a_0^2} \braket{P_X^2\,a_0^2/\hbar^2} },  &&\tilde{\nu}_2 = 4 \sqrt{\braket{X^2/a_0^2}^{ } \braket{p_x^2\,a_0^2/\hbar^2} },  \nonumber \\
\tilde{\nu}_3 &= \sqrt{\braket{y^2/a_0^2}\braket{P_Y^2\,a_0^2/\hbar^2}}, &&
 \tilde{\nu}_4 =4 \sqrt{\braket{Y^2/a_0^2} \braket{p_y^2\,a_0^2/\hbar^2}} , \nonumber \\
\tilde{\nu}_5 &=  \sqrt{\braket{z^2/a_0^2} \braket{P_Z^2\,a_0^2/\hbar^2}},  &&
\tilde{\nu}_6 = 4\sqrt{\braket{Z^2/a_0^2} \braket{p_z^2\,a_0^2/\hbar^2}}. 
\end{align}
Due to the rotational symmetry of the system  in the $x$-$y$ plane, the symplectic eigenvalues are degenerate with  $\tilde{\nu}_1 = \tilde{\nu}_3$ and  $\tilde{\nu}_2 = \tilde{\nu}_4$. The  four unique eigenvalues are  given by 
\begin{align} \label{eq:localised:entanglement:test}
\tilde{\nu}^{(1)}_{nlm} &=\frac{a_0 n }{2b} \sqrt{\frac{\left(l^2+l+m^2-1\right) \left(5 n^2-3 l (l+1) + 1\right)}{4 l^2+4 l-3}}, \nonumber \\
\tilde{\nu}^{(2)}_{nlm} &= 2\sqrt{2} \frac{b}{a_0n} \sqrt{\frac{l^2+l+m^2-1}{ 4 l^2+4 l-3 }}, \nonumber\\
 \tilde{\nu}^{(5)}_{nlm} &=  \frac{a_0 n}{2b} \sqrt{\frac{\left(2 l^2+2l-2 m^2-1\right) \left(5 n^2-3 l (l+1) + 1\right)}{4 l^2+4 l-3}},  \nonumber \\
 \tilde{\nu}^{(6)}_{nlm} &= 2 \sqrt{2} \frac{b}{a_0 n} \sqrt{\frac{2l^2  + 2l - 2m^2 -1}{ 4l^2 + 4l - 3}}. 
\end{align} 
To detect entanglement, we merely require that one of the eigenvalues becomes smaller than unity: ${\nu}^{(j)}_{nlm}< 1$ for $j=1,2, 5,6$. 
The first symplectic eigenvalue $\tilde{\nu}^{(1)}_{nlm}$ indicates  that the likelihood of the PPT criterion being violated increases with $b$, in fact, as $b \rightarrow \infty$, $\tilde{\nu}_j \rightarrow 0$. 

The second eigenvalue $\tilde{\nu}^{(2)}_{nlm}$ has the opposite dependence of $a_0$ and $b$, but scales as $n^{-1}$, which means that the larger $n$ becomes, the more entangled the state appears. We also note that  all eigenvalues depend on the ratio $a_0/b$  (which is independent of the choice of normalisation with either $a_0$ or $b$). 

For the Hydrogenic ground state, with $n = 1$ and $l = m = 0$, we find that  the eigenvalues become even more degenerate with $\tilde{\nu}_{100}^{(5)} = \tilde{\nu}_{100}^{(1)}$ and $\tilde{\nu}_{100}^{(6)} = \tilde{\nu}_{100}^{(2)}$ since rotational symmetry is restored. The remaining unique eigenvalues are
\begin{align} \label{eq:ground:state:symp:eig}
\tilde{\nu}^{(1)}_{100} =  \sqrt{\frac{1}{2}}  \frac{a_0}{b} , \quad \quad \tilde{\nu}^{(2)}_{100} = \sqrt{\frac{8}{3}} \frac{b}{a_0}. 
\end{align}
These values depend purely on the ratio of the Hydrogenic characteristic length-scale and the Gaussian wavepacket spread $a_0/b$. Entanglement can however be detected for most values through examination of these two eigenvalues. When $a_0 \sim b$, the first eigenvalue always detects entanglement. For the special case when $a_0/b \sim \sqrt{2}$, entanglement can briefly not be detected at all, until $a_0/b >\sqrt{8/3}$.  This implies that it would be difficult to verify the entanglement of systems localised to scales that are of the same order as $a_0$.

\begin{figure*}{
  \includegraphics[width=0.5\linewidth, trim = 0mm 0mm 0mm 0mm]{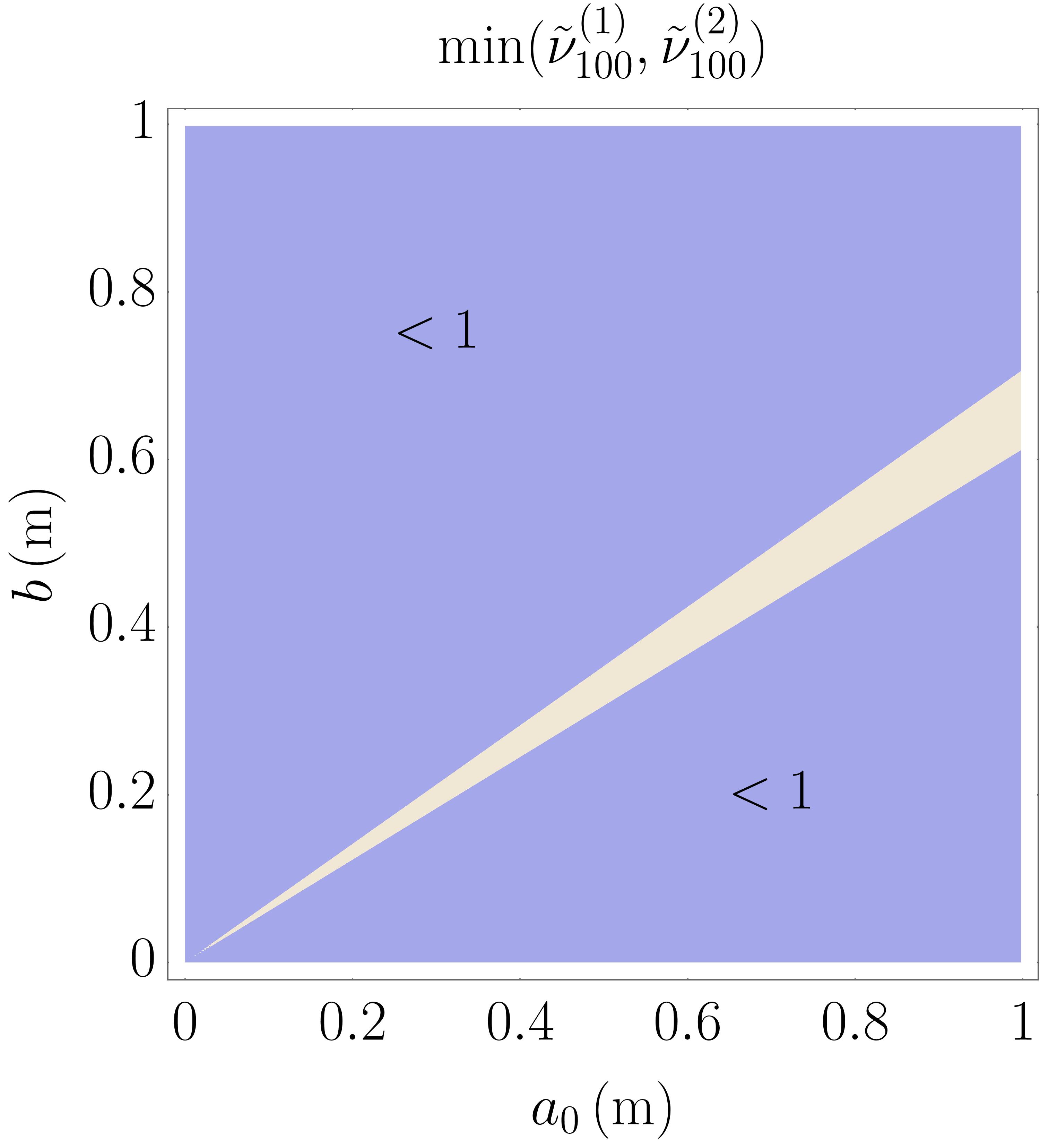}
}  
\caption{\small Plot of  $\mathrm{min}(\tilde{\nu}_{100}^{(1)}, \tilde{\nu}_{100}^{(2)})$  shown in Eq.~\eqref{eq:ground:state:symp:eig} as a function of the  reduced Bohr radius $a_0$ and the Gaussian wavepacket spread $b$.  The plot-range has been chosen merely to demonstrate the validity of the entanglement test. The blue area indicates where either of the two eigenvalues dips below 1, and where entanglement can be detected. The light yellow area indicates where both eigenvalues are larger than 1, and where entanglement cannot be inferred. }
\label{fig:exclusion}
\end{figure*}

We have plotted the function $\mathrm{min}(\tilde{\nu}_{100}^{(1)}, \tilde{\nu}_{100}^{(2)})$ in Figure~\ref{fig:exclusion} to demonstrate the regimes where entanglement can be detected. The blue area indicates where either of the eigenvalues dip below 1, and where entanglement can be inferred. The light yellow area indicates where both eigenvalues are larger than 1,  which means that entanglement cannot be inferred. 

We conclude that in order to apply this entanglement test, one must measure the expectation values and variances of the relative and centre-of-mass coordinates in the $x$ and $z$ spatial directions to compute the two symplectic eigenvalues in  Eq.~\eqref{eq:ground:state:symp:eig}. It is worth mentioning that this test, in contrast to the the previous one based on the relative momentum variance, would be sufficient also for noisy, mixed states (albeit for such states it would likely be more difficult to violate).

\section{Discussion} \label{sec:discussion}
We derived two expressions that can serve as entanglement tests for a free Hydrogenic system (see Eq.~\eqref{eq:Schmidt:rank:free:Hydrogen}) and a localised Hydrogenic system (see Eq.~\eqref{eq:localised:entanglement:test}), respectively. Here, we briefly discuss our results and their potential application to detecting entanglement between two subsystems that interact via a central potential. 

The first additional system that might exhibit Hydrogenic structure is positronium, which is the bound state of one electron and one positron. With a half-life of 0.1244\,ns~\cite{karshenboim2004precision}, it is a highly unstable system. We note that detecting entanglement in positronium might be extremely challenging, however it may also constitute one of the most straight-forward routes towards detecting entanglement between matter and anti-matter systems. 

We speculate that another candidate family of systems that might exhibit Hydrogenic structure could be mesoscopic systems that interact through a central potential, such as a Coulombic potential.  Recent advances in the ability to control the number of charges on levitated silica spheres~\cite{moore2014search} or implant changes through the addition of nitrogen-vacancy (NV) centres~\cite{neukirch2015multi} provide an excellent  means for highly sophisticated operations in the laboratory. It is currently an open question as to whether mesoscopic systems would at all readily exhibit Hydrogenic structure,  but should this be the case, the methods developed here may be used towards establishing any entanglement present between them. 

We further note that any atom with Hydrogenic structure, such as the Helium ion, or any two oppositely charged interacting ions could be treated with our results. Artificial atoms in semiconductor structures are yet another possible setting where tests similar to the ones proposed here might become relevant. In exitonic bound states, one can plausibly have access to each of the entangled components to examine their entanglement. 

Finally, one may also look for entanglement in phenomena beyond the Standard Model. These include  proposals for the existence of so-called magnetic monopoles~\cite{rajantie2012introduction}, which, like opposite electric charges, are theorised to form bound systems of north-south magnetic monopoles. These are referred to as monopolium~\cite{reis2017production}. Should these particles be observed, their interaction might cause the two monopole subsystems to become entangled in a similar fashion to the Hydrogen atom.

\section{Conclusions} \label{sec:conclusions}
In this work, we proposed two entanglement tests for free and localised bipartite systems with Hydrogenic structure. We computed the spread of the Schmidt spectrum for an arbitrary energy eigenstate of a delocalised (free) Hydrogenic system, and showed that it scales with the inverse of the principal quantum number $n^{-1}$. For localised systems, we  demonstrated that measuring the variances of the relative and centre-of-mass coordinates suffices to demonstrate that the system is entangled.  We believe that these results could potentially aid the study of entanglement  in systems that display Hydrogenic characteristics,  such as interacting matter--anti-matter systems or freely-falling mesoscopic systems. 

A number of open questions remain,  such as the extension of these results to dynamical processes and the formation of Hydrogenic states starting from highly localised systems.  One may also ask how the entanglement of a Hydrogenic system changes as a function of an external potential, like, for example, a  magnetic field. The inclusion of an external potentials has previously been shown to drastically change the entanglement contents of related systems~\cite{nazmitdinov2012shape}. We leave  these questions to future work.

\section*{Acknowledgments}
We warmly thank David Cassidy, Michael R. Vanner, Alfred Harwood, Ryan Marshman, Carlo Sparaciari, Alexander Ferrier, Gavin Morley, Peter F. Barker, A. Douglas K. Plato, Dennis R{\"a}tzel, David Edward Bruschi, Alessio Belenchia, Tania Monteiro, Ivette Fuentes, and Markus Aspelmeyer for helpful discussions and comments.  We would also like to convey a special thanks to the referees, whose careful reading of this work was most helpful. SQ is funded by an EPSRC Doctoral Prize Fellowship.

\section*{Bibliography}
\bibliographystyle{iopart-num}
\bibliography{Hydrogen_refs}

\newpage
\appendix

\section{Hydrogenic solutions in position and momentum space} \label{app:Hydrogen:solutions}

In this Appendix, we introduce the Hydrogenic solutions to the time-independent Schr\"{o}dinger equation in three dimensions for two particles that interact through a central potential. 
We describe here the general procedure for separating the Schr{\"o}dinger equation in terms of the centre-of-mass $\vec{R}$ and relative coordinate $\vec{r}$, which will be used to describe the interaction. We begin by presenting the procedure in position space, and then move on to review the corresponding solutions in momentum space. 

\subsection{Hydrogenic wavefunctions in position space}
We follow the standard derivation presented in Appendix 8 of Ref.~\cite{bransden2003physics}. The Schr\"{o}dinger equation for $N$ interacting particles in first quantisation with coordinates $\vec{r}_i$ and masses $m_i $, is given by 
\begin{align}
&- \frac{\hbar^2}{2 } \sum^N_{j = 1} \frac{1}{m_j} \nabla_j^2 \Psi ( \vec{r}_1, \vec{r}_2, \ldots, \vec{r}_N, t) + V (\vec{r}_1 , \vec{r}_2, \ldots, \vec{r}_N, t) \Psi(\vec{r}_1, \vec{r}_2, \ldots \vec{r}_N, t ) \nonumber \\
&\quad\quad\quad\quad=i \hbar \frac{\p}{\p t} \Psi(\vec{r}_1, \vec{r}_2 \ldots \vec{r}_N, t) , 
\end{align}
where $m_j$ is the mass of particle $j$, $\nabla^2_j$ is the Laplacian operator, and  where $V(\vec{r}_1, \vec{r}_2, \ldots, \vec{r}_N, t)$ is the sum of central potentials affecting the particles. For two interacting particles, the equation reduces to
\begin{equation} \label{eq:two-particle-schordinger}
\left[ - \frac{\hbar^2}{2m_1} \nabla^2_1 - \frac{\hbar^2}{2m_2} \nabla^2_2 - V( \vec{r}_1,\vec{r}_2)  \right] \Psi(\vec{r}_1 , \vec{r}_2) = i \hbar \frac{\p }{\p t} \Psi( \vec{r}_1 , \vec{r}_2) ,
\end{equation}
where the explicit form of the potential is given by 
\begin{equation}
V(\vec{r}_1,\vec{r}_2) =  \frac{\alpha}{|\vec{r}_1 - \vec{r}_2|} =   \frac{\alpha}{|\vec{r}|} 
= V(\vec{r}) ,
\end{equation}
where $\alpha >0$ is a coupling constant that depends on the interaction at hand, and $\vec{r} = \vec{r}_1 - \vec{r}_2$. 
For the Coulomb potential, we find that $\alpha = q_1 q_2/(4\pi \epsilon_0)$, and for a Newtonian gravitational interaction, $\alpha = G m_1 m_2$. 

The two-particle Schr\"{o}dinger equation in Eq.~\eqref{eq:two-particle-schordinger} is solved by 
moving to a centre-of-mass coordinate $\vec{R} $ and a relative coordinate $\vec{r}$, defined as 
\begin{align} 
\vec{R} & =  \frac{m_1 \vec{r}_1 + m_2 \vec{r}_2}{ m_1 + m_2} , \label{app:eq:definition:central:coordinates}\\
\vec{r} &=   \vec{r}_1 - \vec{r}_2 ,  \label{app:eq:definition:relative:coordinates}
\end{align}
(with conjugate momenta $\vec{P}=(\vec{p}_1+\vec{p}_2)$ and 
$\vec{p}=(m_2 \vec{p}_1- m_1 \vec{p}_2)/(m_1+m_2)$, defining a canonical quadruple).
We then introduce the reduced mass $\mu = \frac{m_1 m_2}{m_1 + m_2}$ and the total mass $M = m_1 + m_2$ in order to write the original coordinates as
\begin{align}
\vec{r}_1 & = \vec{R} +  \frac{\mu}{m_1} \vec{r}  , \\
\vec{r}_2 &= \vec{R} - \frac{\mu}{m_2} \vec{r}  .
\end{align}
It can then be shown that the gradient operators can be written as
\begin{align}
\boldsymbol{\nabla}_1 &=  \boldsymbol{\nabla}_{\vec{r}} +  \frac{\mu}{m_2} \boldsymbol{\nabla}_{\vec{R}}  ,\\
\boldsymbol{\nabla}_2 &= - \boldsymbol{\nabla}_{\vec{r}} + \frac{\mu}{m_1} \boldsymbol{\nabla}_{\vec{R}}  ,
\end{align} 
which allows us to rewrite the two-particle Schr\"{o}dinger equation as
\begin{equation} \label{eq:two-particle-schrodinger-separated}
\left[  - \frac{\hbar^2}{2(m_1 + m_2)} \nabla^2_{\vec{R}} - \frac{\hbar^2}{2\mu} \nabla_{\vec{r}}^2  - V(\vec{r}) \right] \Psi(\vec{r}, \vec{R}) = E \, \Psi(\vec{r}; \vec{R})  .
\end{equation}
Since the potential only depends on $\vec{r}$ and not on $\vec{R}$, we can separate the variables into the Schr{\"o}dinger equation above, as per $\Psi( \vec{r}, \vec{R} ) = \psi(\vec{r}) \varphi(\vec{R})$, obtaining
\begin{align}
 - \frac{\hbar^2}{2 M } \nabla^2_{\vec{R}} \varphi(\vec{R}) =  E_{\mathrm{CM}} \varphi(\vec{R})   , \label{eq:hydrogen:centre:of:mass:equation}\\
\left[  -\frac{\hbar^2}{2\mu} \nabla_{\vec{r}}^2  + V(\vec{r}) \right] \psi(\vec{r}) = E_{\mathrm{rel}}\psi(\vec{r})  , \label{eq:hydrogen:relative:coordinate:equation}
\end{align}
where $E_{\mathrm{CM}}$ is the centre-of-mass energy and $E_{\mathrm{rel}}$ is the relative energy. The total energy of the system is
\begin{equation}
E = E_{\mathrm{CM}} + E_{\mathrm{rel}} .
\end{equation}
Eq.~\eqref{eq:hydrogen:centre:of:mass:equation} describes the free centre-of-mass wavefunction (the overall wavepacket) of the two-particle system and admits wave-like solutions. 
Eq.~\eqref{eq:hydrogen:relative:coordinate:equation}, on the other hand,  admits the Hydrogenic solutions $\psi_{nlm}(\vec{r})$ with $E_{\mathrm{rel}}$ being the usual Hydrogen energy spectrum~\cite{griffiths2018introduction}. The solutions to the relative equation are given by 
\begin{equation} \label{app:eq:def:Hydrogenic:wavefunctions}
\psi_{nlm}(\vec{r})  = \sqrt{\left( \frac{2}{na_0} \right)^3 \frac{(n-l-1)!}{2n[(n+l)!]^3}}  \, e^{- r/n  a_0} \,  \left( \frac{2r}{na_0} \right)^l \,  \left[ L^{2l+1}_{n-l-1} (2r/na_0) \right] Y^m_l(\theta, \phi)  ,
\end{equation}
where $n,l,m$ are the principal, angular and magnetic quantum numbers, respectively.  $r, \theta$ and $\phi$ are elements of the relative coordinate vector: $\vec{r} = (r, \theta, \phi)^{\mathrm{T}}$, and where $L^{2l+1}_{n-l-1}(2 r/na_0)$ is the associated Laguerre polynomial, defined by
\begin{equation}
L^p_{q-p} (x) \equiv (-1)^p \left( \frac{d}{dx} \right)^p L_q(x)  ,
\end{equation}
and where $L_q(x)$  denotes the so-called Laguerre polynomial, given by 
\begin{equation}
L_q(x) \equiv e^x \left( \frac{d}{dx} \right)^q \left( e^{-x} x^q \right) .
\end{equation}
Finally, we note that $a_0$ is a length scale set by the interaction coefficient $\alpha$, corresponding to the {\em reduced} 
Bohr radius:
\begin{equation} \label{eq:definition:rescaled:Bohr:radius}
a_0 = \frac{\hbar^2}{\mu \, \alpha} .
\end{equation}
Returning to the functions shown  in Eq.~\eqref{app:eq:def:Hydrogenic:wavefunctions}, the spherical harmonics $Y_l^m(\theta, \phi)$  are given by
\begin{equation} \label{app:eq:spherical:harmonics:definition}
Y_l^m(\theta, \phi ) = (-1)^m \sqrt{\frac{(2 l +1)}{4\pi} \frac{(l - m)!}{(l+m)!}} P_{lm}( \cos \theta) \, e^{i m \phi} , 
\end{equation}
where $P_{lm}(\cos \theta)$ are the associated Legendre polynomials without the Condon--Shortley phase~\cite{arfken1999mathematical}. The spherical harmonics satisfy the following orthogonality relation
\begin{equation} \label{app:eq:spherical:harmonics:normalisation}
\int^{\pi}_0 \mathrm{d} \theta  \, \sin \theta  \,\int^{2\pi}_0 \mathrm{d} \phi  \, Y^{m'*}_{l'}(\theta, \phi) Y_l^m(\theta, \phi) = \delta_{ mm'} \delta_{ ll'} .
\end{equation}
We make frequent use of this relation in the following  Appendices. This concludes our summary of the Hydrogenic solutions in position space. 
 
\subsection{Hydrogenic wavefunctions in momentum space}
The Hydrogenic wavefunctions in momentum space $\tilde{\psi}( \vec{k})$, where the relation between the wavevector $\vec{k}$ and the momentum is $\vec{p} = \hbar \vec{k}$, are defined as the Fourier transform of the position space wavefunctions:
\begin{equation}
\tilde{\psi}(\vec{k}) = \int \mathrm{d}\vec{r} \, e^{- i \vec{k} \cdot \vec{r}}  \, \psi(\vec{r}) .
\end{equation}
 Note that we denote this $\vec{k}$ with a bar in the main text to differentiate it from another relative quantity, but we omit this here for notational clarity. The explicit form of the Hydrogenic wavefunctions in momentum space can be computed to be (see Appendix 5 in~\cite{bransden2003physics}):
\begin{equation} \label{app:eq:Hydrogen:wavefunction:momentum:space}
\tilde{\psi}(\vec{k}) = F_{nl} (k)  \, Y_{l}^m(\theta, \phi) ,
\end{equation}
where 
\begin{equation}
F_{nl}(k)= \left[ \frac{2}{\pi} \frac{(n-l-1)!}{(n+l)!} \right]^{1/2} n^2 2^{2l + 2} l! \,  \sqrt{a_0^3} \frac{n^l a_0^l k^l}{(n^2 a_0^2k^2 + 1)^{l+2}} C^{l+1}_{n-l-1} \left( \frac{n^2 a_0^2 k^2 - 1}{n^2 a_0^2 k^2 + 1} \right) ,
\end{equation}
and where $C^\alpha_N(x)$ are the Gegenbauer polynomials~\cite{erdelyi1955higher}, which are defined by the relation
\begin{equation}
(1 - 2 xs + s^2)^{-\alpha}  = \sum^\infty_{N = 0} C^\alpha_N(x) s^N  ,
\end{equation}
with $|s|<1$. The Gegenbauer polynomials are special cases of the more general Jacobi polynomials and play a role that is analogous to the role played by Legendre polynomials in the theory of the three-dimensional spherical harmonics~\cite{avery2012hyperspherical}. 

The Gegenbauer polynomials are normalised by the following orthogonality relation
\begin{equation} \label{eq:Gegenbauer:orthogonality:relation}
\int^1_{-1} \mathrm{d} x \, (1 - x)^{\alpha - 1/2} \,  C_n^{(\alpha)} (x) C^{(\alpha)}_{n'}  = \frac{2^{1 - 2\alpha} \pi  \, \Gamma( n + 2 \alpha) }{n! ( n + \alpha) [\Gamma( \alpha)]^2 }   \delta_{nn'}, 
\end{equation}
where $\alpha$ is a dummy index, not to be mixed up with the coupling constant defined earlier. This is yet another relation that is used many times in what follows.

The normalisation of the momentum wavefunctions reads
\begin{equation}
\int \mathrm{d} \vec{k} \,  |\tilde{\psi}( \vec{k})|^2 = 1, 
\end{equation}
which, since the spherical harmonics are orthogonal (see Eq.~\eqref{app:eq:spherical:harmonics:normalisation}), implies that
\begin{equation} \label{app:eq:momentum:wavefunction:normalisation}
\int^\infty_0 \mathrm{d} k \,  k^2 \, [F_{nl}(k)]^2 = 1.
\end{equation}
This concludes the summary of the Hydrogenic wavefunctions in momentum space. We return to these expressions in~\ref{app:momentum:variance}, where we compute the variance of the relative momentum variable.

\section{Normalisation and expectation values for the momentum Hydrogenic wavefunction} \label{app:momentum:variance}
In this Appendix, we prove  Eq.~\eqref{eq:momentum:variance:result} in the main text, which provides a relation for the variance of the relative momentum coordinate. This, in turn, corresponds to the continuous analogue of the Schmidt coefficients of  a free Hydrogenic system, and is therefore a key result in this work. We begin by proving the orthogonality relation and the normalisation condition of the Gegenbauer polynomials, as this will become useful in proving Eq.~\eqref{eq:momentum:variance:result}. 

 We used the notation $\overbar{\vec{k}}$ in the main text to refer to the local wavevector rather than that related to the relative momentum (which is defined as $\vec{k} = (m_2 \vec{k}_1 - m_1 \vec{k}_2)/(m_1 + m_2)$). We here remove the bar for notational convenience.

\subsection{Normalisation of the Hydrogenic momentum wavefunction}
As discussed in~\ref{app:Hydrogen:solutions}, the momentum representation of the Hydrogenic wavefunctions is given in Eq.~\eqref{app:eq:Hydrogen:wavefunction:momentum:space}. We begin by proving the normalisation relation in Eq.~\eqref{app:eq:momentum:wavefunction:normalisation}, which we reprint here for brevity:
\begin{equation} \label{app:eq:radial:momentum:normalisation}
\int^\infty_0\mathrm{d}k \,  k^2 \,  |F_{nl}(k)|^2  = 1  .
\end{equation}
We reprint the expression for $F_{nl}(k)$ here for convenience:
\begin{equation} \label{app:def:Hydrogen:momentum:wavefunction}
F_{nl}(k)= \left[ \frac{2}{\pi} \frac{(n-l-1)!}{(n+l)!} \right]^{1/2} n^2 2^{2l + 2} l! \,  \sqrt{a_0^3} \frac{n^l a_0^lk^l}{(n^2 a_0^2k^2 + 1)^{l+2}} C^{l+1}_{n-l-1} \left( \frac{n^2 a_0^2k^2  - 1}{n^2 a_0^2k^2  + 1} \right) ,
\end{equation}
We note that momentum wavefunction $\tilde{\psi}(\vec{k})$ has units of length$^{3/2}$ due to the appearance of $\sqrt{a_0^3}$ in Eq.~\eqref{app:def:Hydrogen:momentum:wavefunction}.   This is to ensure that the normalisation is dimensionless.

Our goal is to prove the relation in Eq.~\eqref{app:eq:radial:momentum:normalisation}. We start by writing out the expression in full:
\begin{align}
\int^\infty_0\mathrm{d}k \,  k^2 \,  |F_{nl}(k)|^2= \frac{2}{\pi} &\frac{(n-l-1)!}{(n+l)!} n^4 2^{4l + 4} (l!)^2 \,  a_0^3  \nonumber \\
&\times \int^\infty_0 \mathrm{d} k \,  k^2 \frac{n^{2l} a_0^{2l} k^{2l} }{(n^2 a_0^2k^2 + 1)^{2l+4}} \left[C^{l+1}_{n-l-1} \left( \frac{n^2a_0^2k^2 - 1}{n^2 a_0^2k^2 +1} \right) \right]^2 .
\end{align}
We now multiply and divide by $(a_0 k)^4$ and $4^{l + 2}$ in order to collect the following terms:
\begin{align} \label{app:eq:manipulated:orthogonality:relation}
\int^\infty_0\mathrm{d}k \,  k^2 \,  |F_{nl}(k)|^2
=  \frac{2}{\pi} &\frac{(n-l-1)!}{(n+l)!}  2^{2l } (l!)^2 \, \frac{1}{a_0} \nonumber \\
&\times\int^\infty_0 \mathrm{d}k  \frac{1}{ k^2 } \left( \frac{4 n^2 a_0^2k^2}{(n^2 a_0^2k^2 + 1)^{2}} \right)^{l + 2}\left[C^{l+1}_{n-l-1} \left( \frac{n^2 a_0^2k^2 - 1}{n^2 a_0^2 k^2 +1} \right) \right]^2 .
\end{align}
We now define the new variable
\begin{align} \label{app:eq:variable:sub}
&x = \frac{n^2 a_0^2k^2 - 1}{n^2 a_0^2k^2 + 1}. 
\end{align}
The variable substitution implies that
\begin{align} \label{app:eq:relations}
& k = \frac{1}{a_0n} \sqrt{\frac{1+x}{1-x}}, && \mbox{and} 
&& \frac{4 n^2 a_0^2k^2}{( n^2 a_0^2k^2+1)^2 } = 1 - x^2 , 
\end{align}
and the derivative becomes 
\begin{equation} \label{app:eq:derivative:relation}
\frac{dx}{dk} =  \frac{4 n^2 a_0^2 k}{\left(n^2 a_0^2k^2+1\right)^2} = \frac{1 - x^2}{k}. 
\end{equation}
With this substitution, the limits become $\{-1, 1\}$, which is what we require for the normalisation condition of the Gegenbauer polynomials. 
We insert this into Eq.~\eqref{app:eq:manipulated:orthogonality:relation} to find 
\begin{align} \label{app:eq:momentum:orthogonality:many:manipulations:almost}
\int^\infty_0\mathrm{d}k \,  k^2 \,  |F_{n,l}(k)|^2
&=  \frac{2}{\pi} \frac{(n-l-1)!}{(n+l)!}  2^{2l } (l!)^2 \frac{1}{a_0} \int^1_{-1} \mathrm{d}x \frac{k}{1 - x^2} \frac{1}{ k^2} \left( 1 - x^2\right)^{l+2} \left[C^{l+1}_{n-l-1} \left( x \right) \right]^2 \nonumber \\
&=   \frac{2}{\pi} \frac{(n-l-1)!}{(n+l)!}  2^{2l } (l!)^2 \,n \int^1_{-1} \mathrm{d}x  \,   \sqrt{\frac{1-x}{1+x}}  \left( 1 - x^2\right)^{l+1} \left[C^{l+1}_{n-l-1} \left( x \right) \right]^2 \nonumber \\
&=  \frac{2}{\pi} \frac{(n-l-1)!}{(n+l)!}  2^{2l } (l!)^2\,n  \int^1_{-1} \mathrm{d}x \,   (1-x) (1 - x^2)^{l + 1/2}\left[C^{l+1}_{n-l-1} \left( x \right) \right]^2 .
\end{align}
The last integral in Eq.~\eqref{app:eq:momentum:orthogonality:many:manipulations:almost} can now be written as two separate terms by multiplying out the first bracket:
\begin{align} \label{app:eq:momentum:orthogonality:two:separate:integrals}
\int^\infty_0\mathrm{d}k \,  k^2 \,  |F_{n,l}(k)|^2
&=  \frac{2}{\pi} \frac{(n-l-1)!}{(n+l)!}  2^{2l } (l!)^2  \, n  \int^1_{-1} \mathrm{d}x \,  (1 - x^2)^{l + 1/2}\left[C^{l+1}_{n-l-1} \left( x \right) \right]^2 \nonumber \\
&\qquad- \frac{2}{\pi} \frac{(n-l-1)!}{(n+l)!}  2^{2l } (l!)^2  \, n   \int^1_{-1} \mathrm{d}x \, x(1-x^2)^{l + 1/2} \left[C^{l+1}_{n-l-1} \left( x \right) \right]^2 . 
\end{align}
The integral in the first term of Eq.~\eqref{app:eq:momentum:orthogonality:two:separate:integrals} 
can be evaluated by using the orthogonality relation in Eq.~\eqref{eq:Gegenbauer:orthogonality:relation}. With the index substitution $\alpha \rightarrow l + 1 $ and $n \rightarrow n - l - 1$, we find
\begin{equation} \label{app:eq:Gegenbauer:orthogonality:relation}
\int^1_{-1} \mathrm{d} x \, (1 - x)^{l + 1/2} \,  C_{n - l -1}^{(l + 1)} (x) C^{(l + 1)}_{n - l - 1}(x)  = \frac{2^{-2l - 1} \pi  \, \Gamma( n + l + 1) }{n ( n - l-1)! [\Gamma( l + 1)]^2 }.
\end{equation}
When we include the prefactor, the full first term of Eq.~\eqref{app:eq:momentum:orthogonality:two:separate:integrals} evaluates to
\begin{align} \label{app:eq:momentum:orthogonality:many:manipulations}
 \frac{2}{\pi} \frac{(n-l-1)!}{(n+l)!}  &2^{2l } (l!)^2  \, n\int^1_{-1} \mathrm{d}x \,  (1 - x^2)^{l + 1/2}\left[C^{l+1}_{n-l-1} \left( x \right) \right]^2 \nonumber \\
&= \frac{2}{\pi} \frac{(n-l-1)!}{(n+l)!}  2^{2l } (l!)^2 \, n \frac{2^{-2l - 1} \pi  \, \Gamma( n + l + 1) }{n ( n - l-1)! [\Gamma( l + 1)]^2 }.
\end{align}
Since the Gamma function is given by $\Gamma(n) = (n-1)!$  for a positive integer $n$, we are able to write
\begin{equation}
\frac{2}{\pi} \frac{(n-l-1)!}{(n+l)!}  2^{2l } (l!)^2  \, n \frac{2^{-2l - 1} \pi  \, ( n + l)! }{n ( n - l-1)! (l!)^2 }= 1. 
\end{equation}
This is the normalisation we require. It remains to show that the term on the left-hand side of  Eq.~\eqref{app:eq:momentum:orthogonality:many:manipulations} vanishes. 
As noted in Ref.~\cite{podolsky1929momentum}, to  complete the proof, we can use the following recursion formula for the Gegenbauer polynomials:
\begin{equation}
x \, C^\alpha_n(x) = \frac{1}{2(n + \alpha)} \left[ (n + 1) C^\alpha_{n + 1} (x) + (n + 2\alpha - 1)C^\alpha_{n-1}(x) \right].
\end{equation}
With this relation, the  second integral in Eq.~\eqref{app:eq:momentum:orthogonality:many:manipulations} (without its prefactor and with the appropriate index substitutions) can be written as 
\begin{align}
 \int^1_{-1} \mathrm{d}x \,& x(1-x^2)^{l + 1/2} \left[C^{l+1}_{n-l-1} \left( x \right) \right]^2  \nonumber \\
&= \frac{1}{2n}  \int^1_{-1} \mathrm{d}x  \, (1-x^2)^{l + 1/2} \left[ (n-l) C^{l+1}_{n-l} (x) + (n+l) C^{l+1}_{n-l-2}(x) \right]   C^{l+1}_{n-l-1} \left( x \right) . 
\end{align}
Since the indices of the Gegenbauer polynomials inside and outside the bracket differ, the integral vanishes due to the orthogonality of the Gegenbauer polynomials in Eq.~\eqref{eq:Gegenbauer:orthogonality:relation}. This concludes our proof of Eq.~\eqref{app:eq:radial:momentum:normalisation}.

\subsection{Proof of Hydrogenic wavefunction momentum variance} \label{app:sec:Hydrogen:wavefunction:momentum:variance}
In Section~\ref{sec:free:Hydrogen:general:eigenstates}, we showed that the variance of the relative momentum variable $\vec{k}$ (referred to as $\bar{\vec{k}}$ in the text) corresponds to the continuous analogue of the Schmidt coefficients of the system, which is given by the concise expression in Eq.~\eqref{eq:Schmidt:rank:free:Hydrogen}. Here, we prove the relation in Eq.~\eqref{eq:Schmidt:rank:free:Hydrogen}. 

We wish to show that 
\begin{equation} \label{app:eq:weighted:Gegenbauer:relation}
 \int^\infty_0 \mathrm{d}k \, k^4  \,  [F_{nl}(k)]^2 = \frac{1}{a_0^2n^2} , 
 \end{equation}
where $F_{nl}(k)$ is given in Eq.~\eqref{app:def:Hydrogen:momentum:wavefunction}.
This allows us to write the integral as
\begin{align} \label{app:eq:weighted:start}
 \int^\infty_0 \mathrm{d}k \, k^4  \, [F_{nl}(k)]^2   =  \frac{2}{\pi} &\frac{(n-l-1)!}{(n+l)!} n^4 2^{4l + 4} (l!)^2 \,  a_0^3 \nonumber \\
&\times \int^\infty_0 \mathrm{d}k \, k^4 \, \frac{n^{2l} a_0^{2l}k^{2l}}{(n^2 a_0^2k^2 + 1)^{2l+4}} \left[C^{l + 1}_{n - l -1} \left( \frac{n^2 a_0^2k^2 - 1}{n^2a_0^2k^2+1} \right)\right]^2.
\end{align}
By now multiplying and dividing by $a_0^4$ and $4^{l + 2}$, we can write Eq.~\eqref{app:eq:weighted:start} as 
\begin{align}
 \int^\infty_0 \mathrm{d}k \, k^4  \, F_{nl}(k)F_{nl}^*(k)  =  \frac{2}{\pi} &\frac{(n-l-1)!}{(n+l)!}  2^{2l } (l!)^2 \,  \frac{1}{a_0}  \\ 
&\times \int^\infty_0 \mathrm{d}k  \,\left(  \frac{4n^{2} a_0^{2}k^{2}}{(n^2 a_0^2k^2 + 1)^{2}} \right)^{l + 2} \left[C^{l + 1}_{n - l -1} \left( \frac{n^2 a_0^2k^2 - 1}{n^2a_0^2k^2+1} \right)\right]^2.\nonumber
\end{align}
To evaluate the integral, we perform the same variable substitution as in Eq.~\eqref{app:eq:variable:sub}. Given the  relationships in Eqs.~\eqref{app:eq:relations} and~\eqref{app:eq:derivative:relation}, and the new limits $x \in \{-1, 1\}$, we find
\begin{align} \label{app:eq:weighted:manipulations}
 \int^\infty_0 \mathrm{d}k \, k^4  \, F_{nl}(k)F_{nl}^*(k)
&= \frac{2}{\pi} \frac{(n-l-1)!}{(n+l)!}  2^{2l } (l!)^2 \,  \frac{1}{a_0}\int^1_{-1} \mathrm{d}x \frac{k}{1 - x^2} \left(  1 - x^2 \right)^{l+2} \left[C^{l + 1}_{n - l -1} \left( x\right)\right]^2\nonumber \\
&= \frac{2}{\pi} \frac{(n-l-1)!}{(n+l)!}  2^{2l } (l!)^2 \,  \frac{1}{a_0^2}\int^1_{-1} \mathrm{d}x  \frac{1}{n} \sqrt{\frac{1 + x}{1 - x}} (1 - x^2)^{l+1}\left[C^{l + 1}_{n - l -1} \left( x\right)\right]^2\nonumber \\
&= \frac{2}{\pi} \frac{(n-l-1)!}{(n+l)!}  2^{2l } (l!)^2 \,  \frac{1}{a_0^2 n}\int^1_{-1} \mathrm{d}x \,  (1+x) \left(  1 - x^2 \right)^{l+1/2} \left[C^{l + 1}_{n - l -1} \left( x\right)\right]^2 . 
\end{align}
The last integral in Eq.~\eqref{app:eq:weighted:manipulations} can again be written in terms of two separate terms:
\begin{align}
 \int^\infty_0 \mathrm{d}k \, k^4  \, F_{nl}(k)F_{nl}^*(k)&=   \frac{2}{\pi} \frac{(n-l-1)!}{(n+l)!}  2^{2l } (l!)^2 \,  \frac{1}{a_0^2 n}\int^1_{-1} \mathrm{d}x \left(  1 - x^2 \right)^{l+1/2} \left[C^{l + 1}_{n - l -1} \left( x\right)\right]^2   \\
&\qquad + \frac{2}{\pi} \frac{(n-l-1)!}{(n+l)!}  2^{2l } (l!)^2 \,  \frac{1}{a_0^2 n}\int^1_{-1} \mathrm{d}x \, x \left(  1 - x^2 \right)^{l+1/2} \left[C^{l + 1}_{n - l -1} \left( x\right)\right]^2. \nonumber
\end{align}
As in the previous section, the second term vanishes due to the mismatched indices. Then, by using the orthogonality condition for the Gegenbauer polynomials in Eq.~\eqref{app:eq:Gegenbauer:orthogonality:relation}, we find that the first term evaluates to 
\begin{align}
 \frac{2}{\pi} \frac{(n-l-1)!}{(n+l)!}  2^{2l } (l!)^2 \,  \frac{1}{a_0^2 n}&  \int^1_{-1}  \mathrm{d}x \left(  1 - x^2 \right)^{l+1/2} \left[C^{l + 1}_{n - l -1} \left( x\right)\right]^2 \nonumber \\
&=  \frac{2}{\pi} \frac{(n-l-1)!}{(n+l)!}  2^{2l } (l!)^2 \,  \frac{1}{a_0^2 n}\frac{2^{-2l - 1} \pi  \, \Gamma( n + l + 1) }{n ( n - l-1)! [\Gamma( l + 1)]^2 }.
 \end{align}
Noting that $\Gamma(n) = (n- 1)!$, this expression simplifies to 
\begin{equation}
 \int^\infty_0 \mathrm{d}k \, k^4  \, F_{nl}(k)F_{nl}^*(k) = \frac{2}{\pi} \frac{(n-l-1)!}{(n+l)!}  2^{2l } (l!)^2 \,  \frac{1}{a_0^2 n}\frac{2^{-2l - 1} \pi  \, ( n + l)! }{n ( n - l-1)! (l!)^2 } = \frac{1}{n^2 a_0^2},
\end{equation}
which is the expression we wanted. This concludes the proof of Eq.~\eqref{eq:Schmidt:rank:free:Hydrogen}.

\section{Entropy measures applied to free Hydrogenic systems} \label{app:linear:entropy}

The goal of this Appendix is to compute the linear entropy of a  generic Hydrogenic  energy eigenstate. We provide this result for completeness and we emphasise that it is not possible to invoke an operational interpretation of the linear entropy as a quantifier of the entanglement. 

The continuous variable analogue of the linear entropy $S_{\mathrm{Lin}}$ is 
\begin{equation} \label{app:eq:linear:entropy:position}
S_{\mathrm{Lin}} = 1 - \int \mathrm{d}\vec{r}_1 \int \mathrm{d}\vec{r}_1'   \,\varrho(\vec{r}_1, \vec{r}_1') \, \varrho(\vec{r}_1', \vec{r}_1),
\end{equation}
where $\varrho(\vec{r}_1', \vec{r}_1)$ is the continuous analogue of the subsystem density matrix, given by tracing out one of the variables:
\begin{equation}
\varrho( \vec{r}_1', \vec{r}_1) = \int \mathrm{d} \vec{r}_2  \, \Psi^*(\vec{r}_1', \vec{r}_2) \Psi(\vec{r}_1, \vec{r}_2).
\end{equation}
The form of the linear entropy in Eq.~\eqref{app:eq:linear:entropy:position} is the continuous analogue of the discrete matrix multiplication. 
Inserting the Fourier transform Eq.~\eqref{eq:fourier:transform:inverse}  of the state $\varrho(\vec{r}_1', \vec{r}_1)$  into Eq.~\eqref{app:eq:linear:entropy:position} gives
\begin{align}
S_{\mathrm{Lin}}(\varrho) &=1 -  \int \mathrm{d}\vec{r}_1  \int \mathrm{d}\vec{r}_1'   \,\varrho(\vec{r}_1, \vec{r}_1') \, \varrho(\vec{r}_1', \vec{r}_1) \nonumber \\
&=1 -  \int \mathrm{d}\vec{r}_1 \int \mathrm{d}\vec{r}_1'  \frac{1}{(2\pi)^{3} } \int \mathrm{d}^3 k \,  \tilde{\varrho}(\vec{k}) \,  e^{i \vec{k}\cdot ( \vec{r}_1 - \vec{r}_1')} \frac{1}{(2\pi)^{3}} \int \mathrm{d}^3 k' \, \tilde{\varrho}(\vec{k}') \, e^{i \vec{k}' ( \vec{r}_1' - \vec{r}_1)} \nonumber \\
&= 1- \frac{1}{(2\pi)^{6} } \int \mathrm{d} \vec{k} \,  \tilde{\varrho}(\vec{k})  \int \mathrm{d} \vec{k}' \, \tilde{\varrho}(\vec{k}') \, \int \mathrm{d}\vec{r}_1 \, e^{i \vec{r}_1 ( \vec{k} - \vec{k}')} \int  \mathrm{d}\vec{r}_1'  \, e^{i \vec{r}_1'\cdot ( \vec{k}' - \vec{k})}   \nonumber \\
&=1- \int \mathrm{d} \vec{k} \,  \tilde{\varrho}(\vec{k})    \tilde{\varrho}(\vec{k}) \, \delta(0) \nonumber \\
&=1 -  V\int \mathrm{d} \vec{k} \, [ \tilde{\varrho}(\vec{k}) ]^2, 
\end{align}
where we have defined the Dirac delta function as
\begin{align}
\delta (\vec{k}' - \vec{k}) = \frac{1}{(2\pi)^3}\int \mathrm{d}\vec{x} \, e^{i \vec{x} \cdot (\vec{k}' - \vec{k})}, \quad\quad \mbox{and} \quad\quad \delta(0) =V. 
\end{align}
With the definition of the momentum space Hydrogenic wavefunctions in Eq.~\eqref{app:eq:Hydrogen:wavefunction:momentum:space}, we find
\begin{equation} \label{app:eq:compact:linear:entropy}
S_{\mathrm{Lin}} = 1 -  \frac{1}{V} \int \mathrm{d} \vec{k} \left[ F_{nl} (k) \right]^4 | Y_{l}^m(\theta, \phi) |^4 . 
\end{equation}
To the authors' knowledge, this integral has no known standard solutions. Instead, we aim to derive a closed-form expression that can be evaluated for specific choices of $n$, $l$ and $m$. 
To simplify the evaluation of the integral, we write it in terms of a radial integral $I_k$ and an angular integral $I_\theta$, such that 
\begin{equation}
S_{\mathrm{Lin}} = 1 -\frac{1}{V} I_{\mathrm{rad}} I_{\mathrm{ang}}. 
\end{equation}
We now proceed to evaluate each integral separately. 

\vspace{0.8cm}

\subsection{The angular integral}
Let us start by evaluating the angular integral, which involves four spherical harmonics:
\begin{equation}
I_{\mathrm{ang}} = \int^{\pi}_0 \mathrm{d} \theta \, \sin \theta \int^{2\pi}_0 \mathrm{d}\phi  \, |Y_l^m(\theta, \phi)|^4 . 
\end{equation}
To evaluate this integral, we first seek to reduce the order of the spherical harmonics from four to two. We use the following Clebsch--Gordon expansion~\cite{inghoff2001maple} to write
\begin{align} \label{app:linear:entropy:Y:expansion}
Y_{l_1, m_1}& ( \theta , \phi) \, Y_{l_2, m_2}(\theta, \phi)  \\
 &= \sqrt{\frac{(2l_1 +1)(2l_2 + 1)}{4\pi}}  \sum_{l_3, m_3} ( - 1)^{m_3} \sqrt{2 l_3 + 1} 
\begin{pmatrix} 
l_1 & l_2 & l_3 \\
m_1 & m_2 & - m_3 \end{pmatrix}
\begin{pmatrix} 
l_1 & l_2 & l_3 \\
0 & 0 & 0 \end{pmatrix} 
Y_{l_3, m_3}(\theta, \phi), \nonumber
\end{align}
where the $3\times2$ matrices are in fact scalar functions known as the Wigner 3-$j$ symbols. They are defined in terms of the Clebsch--Gordon coefficients and read:
\begin{equation}
\begin{pmatrix}
j_1 & j_2 & j_3 \\
m_1 & m_2 & m_3 \end{pmatrix} \equiv \frac{(-1)^{j_1 - j_2-m_3}}{\sqrt{2j_3 + 1}} \braket{j_1 m_1 j_2 m_3 | j_3(-m_3)}, 
\end{equation}
where $\braket{j_1 m_1 j_2 m_3 | j_3(-m_3)}$ denotes the inner product between two eigenstates. The Wigner 3-$j$ are zero unless the following conditions are fulfilled:
\begin{align}
&m_i \in \{ - j_i , -j_i +1, - j_i + 2 , \ldots, j_i \}, \quad \mbox{for} \quad i = 1,2,3, \nonumber \\
&m_1 + m_2 + m_3 = 0,  \nonumber \\
&|j_1 -j_2| \leq j_3 \leq j_1 + j_2,  \nonumber \\
& j_1 + j_2 + j_3 \mbox{ is an integer}. 
\end{align}
When $l_1 = l_2 = l$ and $m_1 = m_2 = m$, as in our case, we find
\begin{align}
Y_{l m} ( \theta , \varphi) \, Y_{lm}(\theta, \varphi) &= \frac{(2l +1)}{2\sqrt{\pi}} \sum_{l', m'}^{2l} ( - 1)^{m'} \sqrt{2 l' + 1} 
\begin{pmatrix} 
l & l & l' \\
m & m & - m'\end{pmatrix}
\begin{pmatrix} 
l & l & l' \\
0 & 0 & 0 \end{pmatrix} 
Y_{l', m'}(\theta, \varphi) \nonumber \\
&\equiv \sum_{l', m'} f_{l, m,l', m'} Y_{l' m'}(\theta, \varphi), 
\end{align}
where $f_{l, m, l', m'}$ is defined as 
\begin{equation} \label{app:eq:definition:flmlm}
f_{l, m, l', m'} =  ( - 1)^{m'} \frac{(2l +1 )\sqrt{2 l' + 1}}{2 \sqrt{\pi}} 
\begin{pmatrix} 
l & l & l' \\
m & m & - m' \end{pmatrix}
\begin{pmatrix} 
l & l & l' \\
0 & 0 & 0 \end{pmatrix}. 
\end{equation}
Transforming two of the spherical harmonics in Eq.~\eqref{app:eq:compact:linear:entropy} allows us to then use the normalisation condition in Eq.~\eqref{app:eq:spherical:harmonics:normalisation} to write: 
\begin{align}
I_{\mathrm{ang}} =&\int^\pi_0  \mathrm{d} \theta \, \sin{\theta}\, \int^{2\pi}_0 \mathrm{d}\phi \,  Y_{l}^m(\theta,\phi)   \,  Y_{l}^m(\theta,\phi) \, Y_{l}^{m*} (\theta, \phi)  \, Y_{l}^{m*} (\theta, \phi)\nonumber  \\
&= \int^\pi_0  \mathrm{d} \theta \, \sin{\theta} \, \int^{2\pi}_0  \mathrm{d}\phi \, \sum_{l', m'} \, f_{l, m,  l', m'}  Y_{l'}^{m'}(\theta, \phi) \, \sum_{l'', m''}\, f_{l, m,l'', m''}^* Y_{l''}^{ m''*}(\theta, \phi)\nonumber  \\
& = \sum_{l', m', l'', m''} \,f_{l, m, l', m'} \,f_{l, m, l'', m''}^* \delta_{l', l''} \delta_{m', m''} \nonumber \\
&=   \sum_{l',m'} \, \,f_{l,m,l', m'} \,f_{l, m, l', m'}^*. 
\end{align}
Using the definition of $f_{l, m, l', m'}$ in Eq.~\eqref{app:eq:definition:flmlm}, we expand the expression to find
\begin{align}
I_{\mathrm{ang}}=&    \sum_{l',m'} \, \,f_{l, m, l', m'} \,f_{l, m,  l', m'}^* \\
=& \sum_{l',m'} \frac{(2l +1)^2(2 l' + 1)}{4\pi} 
\begin{pmatrix} 
l & l & l' \\
m & m & - m' \end{pmatrix}
\begin{pmatrix} 
l & l & l' \\
0 & 0 & 0 \end{pmatrix} 
\begin{pmatrix} 
l & l& l' \\
m& m& - m' \end{pmatrix}^*
\begin{pmatrix} 
l & l & l' \\
0 & 0 & 0 \end{pmatrix}^*  . \nonumber
\end{align}
One of the requirements to ensure that the Wigner 3-$j$ symbols are non-zero states that the second row satisfies $m_1 + m_2 + m_3 = 0$. In our case, this corresponds to $2m = m'$. This requirement is only fulfilled for a single value of the sum, namely $m' = 2m$, which cancels the sum over $m'$. Hence we obtain
\begin{align} \label{app:eq:angular:integral}
I_{\mathrm{ang}}&= \sum_{l'} \frac{(2l +1)^2(2 l' + 1)}{4\pi} 
\begin{pmatrix} 
l & l & l' \\
m & m & -2m\end{pmatrix}^2
\begin{pmatrix} 
l & l & l' \\
0 & 0 & 0 \end{pmatrix} ^2. 
\end{align}
We proceed to compute the radial integral. 

\subsection{The radial integral}
The radial integral is given by 
\begin{equation} \label{app:linear:entropy:radial:integral:definition}
I_{\mathrm{rad}} = \int^\infty_0 \mathrm{d} k \,  k^2 \,  |F_{nl}(k)|^4. 
\end{equation}
Now,  $F_{nl}(k)$ is given in Eq.~\eqref{app:def:Hydrogen:momentum:wavefunction}. Just like we did in~\ref{app:momentum:variance}, 
we rewrite the integral in Eq.~\eqref{app:linear:entropy:radial:integral:definition} by dividing and multiplying with $a_0^8 k^6$ and $4^{2l + 4}$ to find: 
\begin{align}
I_{\mathrm{rad}} &= \left[ \frac{2}{\pi} \frac{(n-l-1)!}{(n+l)!} \right]^{2} n^8 2^{8l + 8} (l!)^4 a_0^6 \int^\infty_0 \mathrm{d} k \, k^2  \frac{n^{4l} a_0^{4l} k^{4l}}{(n^2 a_0^2 k^2 + 1)^{4l+8}} \left[ C^{l+1}_{n-l-1} \left( \frac{n^2 a_0^2 k^2 - 1}{n^2 a_0^2 k^2 + 1} \right)\right]^4 \nonumber \\
&=  \left[ \frac{2}{\pi} \frac{(n-l-1)!}{(n+l)!} \right]^{2}  2^{2l } (l!)^4\frac{1}{a_0^2} \int^\infty_0 \mathrm{d} k \,  \frac{1}{ k^6}  \left( 
\frac{4n^{2}a_0^2k^{2}}{(n^2 a_0^2k^2 + 1)^{2}}\right)^{2l+4} \left[ C^{l+1}_{n-l-1} \left( \frac{n^2a_0^2 k^2 - 1}{n^2 a_0^2k^2 + 1} \right)\right]^4. 
\end{align}
We again perform the same variable substitution as in Eq.~\eqref{app:eq:variable:sub} to find:
\begin{align}
I_ {\mathrm{rad}} &=\left[ \frac{2}{\pi} \frac{(n-l-1)!}{(n+l)!} \right]^{2}  2^{2l} (l!)^4 \frac{1}{a_0^2}\int^\infty_0 \mathrm{d} k \,  \frac{1}{ k^6}  \left( 
\frac{n^{2} a_0^2 k^{2}}{(n^2 a_0^2 k^2 + 1)^{2}}\right)^{2l+4} \left[ C^{l+1}_{n-l-1} \left( \frac{n^2 a_0^2 k^2 - 1}{n^2 a_0^2 k^2 + 1} \right)\right]^4 \nonumber \\
&= \left[ \frac{2}{\pi} \frac{(n-l-1)!}{(n+l)!} \right]^{2}  2^{2l} (l!)^4\frac{1}{a_0^2} \int^1_{-1} \mathrm{d} x \, \frac{1}{k^6} \frac{k}{1 - x^2}  \left( 
1 - x^2\right)^{2l+4} \left[ C^{l+1}_{n-l-1} \left(x \right)\right]^4 \nonumber \\
&=  \left[ \frac{2}{\pi} \frac{(n-l-1)!}{(n+l)!} \right]^{2} 2^{2l } (l!)^4  n^5 a_0^3  \int^1_{-1} \mathrm{d} x \,   (1 - x)^{5}  \left( 
1 - x^2\right)^{2l+1/2} \left[ C^{l+1}_{n-l-1} \left(x \right)\right]^4 . 
\end{align}
This integral does not satisfy the Gegenbauer normalisation condition in Eq.~\eqref{eq:Gegenbauer:orthogonality:relation}. We must instead find another way to evaluate the integral. To this end, there exists an alternative expansion of the Gegenbauer polynomials in terms of polynomials, given by~\cite{san2012some}
\begin{equation}
C_n^{(\lambda)} (x) = \begin{pmatrix} n + 2 \lambda - 1 \\ n \end{pmatrix} \sum_{k = 0}^n \begin{pmatrix} n \\k \end{pmatrix} \frac{(2 \lambda + n )_k }{(\lambda + 1/2)_k} \left( \frac{x- 1}{2} \right)^k = \sum_{k = 0}^n f_{ k} \left( 1 - x\right)^k , 
\end{equation}
where we have defined
\begin{equation}
f_{ k } = \begin{pmatrix} n + 2 \lambda - 1 \\ n \end{pmatrix}\begin{pmatrix} n \\k \end{pmatrix} \frac{(2 \lambda + n )_k }{(\lambda + 1/2)_k} (-1)^k  \frac{1}{2^k}, 
\end{equation}
and where $a_k$ is the Pochhammer symbol, which is defined such that $a_k = a(a+1) (a+2) \ldots (a + k - 1)$. Inserting this expansion into the integral, we find
\begin{align}
I_r &=  \left[ \frac{2}{\pi} \frac{(n-l-1)!}{(n+l)!} \right]^{2}  2^{2l} (l!)^4 n^5 a_0^3  \int^1_{-1} \mathrm{d} x \,    (1 - x)^{5}  \left( 1 - x^2\right)^{2l+1/2} \left[ C^{l+1}_{n-l-1} \left(x \right)\right]^4   \\
&=  \left[ \frac{2}{\pi} \frac{(n-l-1)!}{(n+l)!} \right]^{2} 2^{2l } (l!)^4 n^5 a_0^3\sum_{a,b,c,d}  f_{ a} f_{b}f_{c} f_{d} 
 \int^1_{-1} \mathrm{d} x \,  (1 - x)^{5+a + b+ c + d}\left( 1 - x^2\right)^{2l+1/2} . \nonumber
\end{align}
Defining $  \gamma =a + b + c + d$, we find 
\begin{align}
 \int^1_{-1} \mathrm{d} x \,  (1 - x)^{5+\gamma}\left( 1 - x^2\right)^{2l+1/2}  &= \frac{(\gamma+5) \, _3F_2\left(1,-\frac{\gamma}{2}-2,-\frac{\gamma}{2}-\frac{3}{2};\frac{3}{2},2 l+\frac{5}{2};1\right)}{4 l+3} \nonumber \\
&\quad +\frac{\, _3F_2\left(\frac{1}{2},1,-2 l-\frac{1}{2};\frac{\gamma}{2}+\frac{7}{2},\frac{\gamma}{2}+4;1\right)}{\gamma+6} \nonumber \\
&\quad +\frac{2^{4 l+\gamma+6} \Gamma \left(2 l+\frac{3}{2}\right) \Gamma \left(2 l+\gamma+\frac{13}{2}\right)}{\Gamma (4 l+\gamma+8)}, 
\end{align}
where $\,_3F_2$ is the generalised hypergeometric function. Together with its prefactor, the radial integral becomes
\begin{align} \label{app:eq:radial:integral}
I_{\mathrm{rad}} =& \,  \left[ \frac{2}{\pi} \frac{(n-l-1)!}{(n+l)!} \right]^{2}(l!)^4 2^{2l} n^5 a_0^3\left[ \begin{pmatrix} n +  l  \\ n-l-1\end{pmatrix}\right]^4 \nonumber \\
&\times \sum_{a, b, c, d} \frac{(-1)^{a+b+c+d}}{2^{a + b + c+ d}} 
\begin{pmatrix} n-l-1 \\ a \end{pmatrix} 
\frac{(n + l + 1)_a }{(l+ 3/2)_a}  
\begin{pmatrix} n-l-1 \\ b \end{pmatrix}  \frac{(n + l + 1)_b }{(l+ 3/2)_b}   \nonumber \\
&\quad\quad\quad \times  
\begin{pmatrix} n-l-1 \\ c \end{pmatrix} 
\frac{(n + l + 1 )_c }{(l+3/2)_c}    
\begin{pmatrix} n-l-1 \\ d \end{pmatrix}
\frac{(n + l + 1)_d}{(l+3/2)_d}  \nonumber \\
&\times\biggl[ \frac{2^{4 l+a + b + c + d+6} \Gamma \left(2 l+\frac{3}{2}\right) \Gamma \left(2 l+a + b + c + d+\frac{13}{2}\right)}{\Gamma (4 l+a + b + c + d+8)} \nonumber \\
&\quad\quad +\frac{\, _3F_2\left(\frac{1}{2},1,-2 l-\frac{1}{2};\frac{a + b + c + d}{2}+\frac{7}{2},\frac{a + b + c + d}{2}+4;1\right)}{a + b + c + d+6} \nonumber \\
&\quad\quad+(a + b+ c + d+5) \frac{  _3F_2\left(1,-\frac{a + b+ c + d}{2}-2,-\frac{a + b+ c + d}{2}-\frac{3}{2};\frac{3}{2},2 l+\frac{5}{2};1\right)}{4 l+3}  \biggr].
\end{align}
We are now ready to put everything together. 

\subsection{Final expression for the linear entropy}
Combining the angular integral in Eq.~\eqref{app:eq:angular:integral} and the radial integral in Eq.~\eqref{app:eq:radial:integral}, we find
\begin{align} \label{app:eq:final:expression:linear:entropy}
S_{\mathrm{Lin}} =& \, 1 -  \frac{1}{ V} I_{\mathrm{rad}} I_{\mathrm{ang}} \nonumber \\
=& \, 1-  \frac{n^5 a_0^3}{ V} \sum_{l'} \frac{(2l +1)^2(2 l' + 1)}{4\pi} 
\begin{pmatrix} 
l & l & l' \\
m & m & -2m\end{pmatrix}^2
\begin{pmatrix} 
l & l & l' \\
0 & 0 & 0 \end{pmatrix}^2 \nonumber \\
&\times  \left[ \frac{2}{\pi} \frac{(n-l-1)!}{(n+l)!} \right]^{2}(l!)^4 2^{2l} \left[ \begin{pmatrix} n +  l  \\ n-l-1\end{pmatrix}\right]^4 \nonumber \\
&\times \sum_{a, b, c, d} \frac{(-1)^{a+b+c+d}}{2^{a + b + c+ d}} 
\begin{pmatrix} n-l-1 \\ a \end{pmatrix} 
\frac{(n + l + 1)_a }{(l+ 3/2)_a}  
\begin{pmatrix} n-l-1 \\ b \end{pmatrix}  \frac{(n + l + 1)_b }{(l+ 3/2)_b}   \nonumber \\
&\quad\quad\quad \times  
\begin{pmatrix} n-l-1 \\ c \end{pmatrix} 
\frac{(n + l + 1 )_c }{(l+3/2)_c}    
\begin{pmatrix} n-l-1 \\ d \end{pmatrix}
\frac{(n + l + 1)_d}{(l+3/2)_d}  \nonumber \\
&\times\biggl[ \frac{2^{4 l+a + b + c + d+6} \Gamma \left(2 l+\frac{3}{2}\right) \Gamma \left(2 l+a + b + c + d+\frac{13}{2}\right)}{\Gamma (4 l+a + b + c + d+8)}\nonumber \\
&\quad\quad+\frac{\, _3F_2\left(\frac{1}{2},1,-2 l-\frac{1}{2};\frac{a + b + c + d}{2}+\frac{7}{2},\frac{a + b + c + d}{2}+4;1\right)}{a + b + c + d+6} \nonumber \\
&\quad\quad+(a + b+ c + d+5) \frac{  _3F_2\left(1,-\frac{a + b+ c + d}{2}-2,-\frac{a + b+ c + d}{2}-\frac{3}{2};\frac{3}{2},2 l+\frac{5}{2};1\right)}{4 l+3}  \biggr].
\end{align}
This expression is not particularly elegant (perhaps there are also other, more suitable ways to evaluate the radial integral --  the present authors were not able to find a way to do so), but it can be evaluated for specific choices of $n$,  $l$ and $m$. For the Hydrogenic ground state, for example, the linear entropy becomes
\begin{equation}
S_{\mathrm{Lin}} =  1-  \frac{1}{V} \frac{33 \, a_0^3}{16 \, \pi^2}, 
\end{equation}
where we note that the fraction is dimensionless due to the appearances of  the integration volume $V$ and  the reduced Bohr radius $a_0$. 
Since we take $V\rightarrow \infty$ at the end of each calculation, we are left with $S_{\mathrm{Lin}} = 1$, which conventionally implies that the entropy is infinite. However, as mentioned, the linear entropy cannot be interpreted as a measure of entanglement in this context. A similar result for the ground state linear entropy of the Hydrogen atom was reported in a Masters thesis, see Ref.~\cite{graminvestigation}.

\section{Covariance matrix and symplectic eigenvalues} \label{app:sec:covariance:matrix}
In the main text, we noted that successfully detecting entanglement in the first and second moments of a Gaussian state through the application of the PPT criterion immediately implies that a non-Gaussian state with the same moments is also entangled~\cite{serafini2017quantum}. In this Appendix, we demonstrate in detail how we compute the PPT criterion for Hydrogenic systems, by first performing a symplectic transformation from the $\{\vec{r}_1, \vec{r}_2\}$-basis into the relative $\{\vec{r}, \vec{R}\}$ basis. We provide the explicit form of the symplectic transformation and compute the symplectic eigenvalues of the final covariance matrix. 

Our task is to compute the entanglement between the two subsystems  in the $\{\vec{r}_1, \vec{r}_2\}$ basis. We define $\sigma$ in this basis with elements given by
\begin{equation}
\sigma = \braket{\mathbb{X}_i \mathbb{X}_j + \mathbb{X}_j \mathbb{X}_j} - \braket{\mathbb{X}_i} \braket{\mathbb{X}_j}. 
\end{equation}
The diagonal elements of $\sigma$ are given by 
\begin{equation}
\mathrm{diag}[\sigma] =2 \,  \left( \braket{x_1^2} , \braket{p_{x, 1}^2} , \braket{y_1^2} , \braket{p_{y, 1}^2 } , \braket{z_1^2},  \braket{p_{z, 1}^2}, \braket{x_2^2}, \braket{p_{x,2}^2} , \braket{y_2^2}, \braket{p_{y, 2}^2} , \braket{z_2^2}, \braket{p_{z,2}^2} \right). 
\end{equation}
Where $x_j, y_j, z_j$ and $p_{x,j}, p_{y, j}, p_{z,j}$ are the position and momentum coordinates of the two respective systems. 

To transform into the basis of the subsystems parametrised by $\vec{r}_1$ and $\vec{r}_2$, we perform a symplectic transformation $S$ that maps the covariance matrix from the $\{\vec{r}_1, \vec{p}_1, \vec{r}_2,
\vec{p}_2\}$ basis to the $\{\vec{r}, \vec{p}, \vec{R}, \vec{P}\}$ basis.

Assuming that the two masses of the systems are the same, the symplectic transformation $S$ that maps the covariance matrix from the relative and centre-of-mass basis to the subsystem basis is given by 
\begin{equation} \label{app:eq:symplectic:transformation}
S = 
\left( 
\begin{array}{cccccccccccc}
 1 & 0 & 0 & 0 & 0 & 0 & -1 & 0 & 0 & 0 & 0 & 0 \\
 0 & \frac12 & 0 & 0 & 0 & 0 & 0 & -\frac12 & 0 & 0 & 0 & 0 \\
 0 & 0 & 1 & 0 & 0 & 0 & 0 & 0 &-1 & 0 & 0 & 0 \\
 0 & 0 & 0 & \frac12 & 0 & 0 & 0 & 0 & 0 & -\frac12& 0 & 0 \\
 0 & 0 & 0 & 0 & 1 & 0 & 0 & 0 & 0 & 0 & -1& 0 \\
 0 & 0 & 0 & 0 & 0 & \frac12 & 0 & 0 & 0 & 0 & 0 & -\frac12\mbox{ } \\
\mbox{ }\frac12\mbox{ } & 0 & 0 & 0 & 0 & 0 & \frac12 & 0 & 0 & 0 & 0 & 0 \\
 0 & \mbox{ }1\mbox{ } & 0 & 0 & 0 & 0 & 0 & 1 & 0 & 0 & 0 & 0 \\
 0 & 0 & \mbox{ }\frac12\mbox{ } & 0 & 0 & 0 & 0 & 0 & \frac12 & 0 & 0 & 0 \\
 0 & 0 & 0 & \mbox{ }1\mbox{ } & 0 & 0 & 0 & 0 & 0 & 1 & 0 & 0 \\
 0 & 0 & 0 & 0 & \mbox{ }\frac12\mbox{ } & 0 & 0 & 0 & 0 & 0 & \frac12 & 0 \\
 0 & 0 & 0 & 0 & 0 & \mbox{ }1\mbox{ } & 0 & 0 & 0 & 0 & 0 & 1 \\
\end{array}
\right). 
\end{equation}
In language commonly used in the context of quantum optics, this transformation is equivalent to the combination of a 50:50 beam-splitter 
and a diagonal squeezer. 

The new covariance matrix is given by $\sigma' = S \sigma S^{\mathrm{T}}$. We prove in the following Appendix that all the first moments in the relative basis are zero, and that $\sigma'$ is a diagonal matrix with the following entries:
\begin{align} 
\sigma' =2 \, \mathrm{diag} \left( \braket{x^2}, \braket{p_x^2}, \braket{y^2}, \braket{p_y^2}, \braket{z^2}, \braket{p_z^2}, \braket{X^2}, \braket{P_X^2}, \braket{Y^2}, \braket{P_Y^2}, \braket{Z^2}, \braket{P_Z^2} \right), 
\end{align}
where we have defined the relative and centre-of-mass position coordinates as follows:
\begin{align} \label{app:eq:redefined:position:coordinates}
 &x= {x_1 - x_2}, &&X = \frac{x_1 + x_2}{2}, \nonumber \\
&y = {y_1 - y_2}, &&Y = \frac{y_1 + y_2}{2}, \nonumber\\
&z = {z_1 - z_2}, &&Z = \frac{z_1 + z_2}{2}, 
\end{align}
and, similarly, the momentum variables:
\begin{align} \label{app:eq:redefined:momentum:coordinates}
 &p_x= \frac{p_{x,1} - p_{x,2}}{2}, &&P_X = p_{x,1} + p_{x,2}, \nonumber \\
&p_y = \frac{p_{y,1} - p_{y,2}}{2}, &&P_Y = p_{y,1} + p_{y,2}, \nonumber\\
&p_z = \frac{p_{z,1} - p_{z,2}}{2}, &&P_Z = p_{z,1} + p_{z,2}, 
\end{align}
Once the expectation values of the coordinates in  Eqs.~\eqref{app:eq:redefined:position:coordinates} and~\eqref{app:eq:redefined:momentum:coordinates} have been computed, we transform back into the original basis with $\sigma = S^{-1} \sigma' S^{-1\,\mathrm{T}}$.
To better see the explicit action of this transformation and the return of coherence to the off-diagonal elements, we write $\sigma$ in terms of block matrices:
\begin{equation}
\sigma =\left(
\begin{array}{c|c}
\sigma_1 & \sigma_{12} \\
\hline
\sigma_{12} & \sigma_2
\end{array}
\right) .
\end{equation}
Then, since the subsystems are symmetric, with $ \varrho_1( \vec{r}_1', \vec{r}_1) = \varrho_2( \vec{r}_1', \vec{r}_1)$, the covariance matrices for both systems are the same with $\sigma_1 = \sigma_2$, where $\sigma_1$ is given by
\arraycolsep=0.1pt\def\arraystretch{1.2}
{\small
\begin{equation} \label{eq:sigma1:subsystem}
\sigma_1 =  2 \left(
\begin{array}{cccccc}
 \frac14\braket{x^2}+\braket{X^2} & 0 & 0 & 0 & 0 & 0 \\
 0 & \braket{p_x^2}+\frac14\braket{P_X^2} & 0 & 0 & 0 & 0 \\
 0 & 0 & \frac14\braket{y^2}+\braket{Y^2} & 0 & 0 & 0 \\
 0 & 0 & 0 & \braket{p_y^2}+\frac14\braket{P_Y^2} & 0 & 0 \\
 0 & 0 & 0 & 0 & \frac14\braket{z^2}+\braket{Z^2} & 0 \\
 0 & 0 & 0 & 0 & 0 & \braket{p_z^2}+\frac14\braket{P_Z^2} \\
\end{array}
\right).
\end{equation}
}
The off-diagonal elements, which contain the correlations between subsystem 1 and 2, are given by 
{\small
\begin{equation}
\sigma_{12} = 2\left(
\begin{array}{cccccc}
\braket{X^2}-\frac14\braket{x^2} & 0 & 0 & 0 & 0 & 0 \\
 0 &\frac14\braket{P_X^2}-\braket{p_x^2} & 0 & 0 & 0 & 0 \\
 0 & 0 & \braket{Y^2}-\frac14\braket{y^2} & 0 & 0 & 0 \\
 0 & 0 & 0 &\frac14\braket{P_Y^2}-\braket{p_y^2} & 0 & 0 \\
 0 & 0 & 0 & 0 & \braket{Z^2}-\frac14\braket{z^2} & 0 \\
 0 & 0 & 0 & 0 & 0 & \frac14 \braket{P_Z^2}-\braket{p_z^2} \\
\end{array}
\right).
\end{equation}}
We are now ready to perform the partial transposition of $\sigma$ and compute its symplectic eigenvalues to determine whether the $\vec{r}_1$ and $\vec{r}_2$ subsystems are entangled. We call $\nu_j$ the symplectic eigenvalues of $\sigma$ and $\tilde{\nu}_j$ the symplectic eigenvalues of the partially transposed covariance matrix $\sigma^{\mathrm{Tp}}$. 

In a continuous variable system, partial transposition is equivalent to introducing a minus sign to all the momenta (or all the positions) of a subsystem~\cite{serafini2017quantum}. This implies that every element corresponding to the relative momentum expectation values $\braket{p_{x,2}}, \braket{p_{y, 2}}, \braket{p_{z,2}}$ becomes inverted. The diagonal elements of $\sigma^{\prime \mathrm{Tp}}$ remain invariant, and the off-diagonal elements $\tilde{\sigma}_{12}$ are now given by 
{\small
\begin{equation}
\tilde{\sigma}_{12} = 2\left(
\begin{array}{cccccc}
 \braket{X^2}-\frac{1}{4}\braket{x^2} & 0 & 0 & 0 & 0 & 0 \\
 0 &\braket{p_x^2}-\frac14\braket{P_X^2} & 0 & 0 & 0 & 0 \\
 0 & 0 & \braket{Y^2}-\frac14\braket{y^2} & 0 & 0 & 0 \\
 0 & 0 & 0 &\braket{p_y^2}-\frac14\braket{P_Y^2} & 0 & 0 \\
 0 & 0 & 0 & 0 & \braket{Z^2}-\frac14\braket{z^2} & 0 \\
 0 & 0 & 0 & 0 & 0 & \braket{p_z^2}-\frac14 \braket{P_Z^2} \\
\end{array}
\right).
\end{equation}
}
We can now compute the symplectic eigenvalues of the partially transposed system. They are given by the eigenvalues of the product $ i \Omega \sigma^{\prime \mathrm{Tp}}$, and we find:
\begin{align} \label{eq:symplectic:eigenvalues}
\tilde{\nu}_1 &=   \sqrt{   \braket{x^2} \braket{P_X^2} } , 
&&\tilde{\nu}_2 = 4\, \sqrt{\braket{X^2} \braket{p_x^2} },  \nonumber \\
\tilde{\nu}_3 &= \sqrt{\braket{y^2}\braket{P_Y^2}},  &&
 \tilde{\nu}_4 =4\, \sqrt{\braket{Y^2} \braket{p_y^2}}, \nonumber \\
\tilde{\nu}_5 &= \sqrt{\braket{z^2} \braket{P_Z^2}},  &&
\tilde{\nu}_6 = 4 \,\sqrt{\braket{Z^2} \braket{p_z^2}}.
\end{align}
We note that the symplectic eigenvalues do not mix between the modes. Entanglement can then be established once the first and second moments of the Hydrogenic system have been computed, which we do in the next Appendix. 
Due to the rotational symmetry of the system, these three pairs of quantities will take the same along each spatial direction, so that 
our entanglement test will effectively reduce to a two-mode problem.

\section{First and second moments of a localised Hydrogenic system} \label{app:localised:Hydrogen}
In this Appendix, we compute the first and second moments of the Hydrogenic energy eigenstates for the case when the centre-of-mass wavefunction is a three-dimensional localised Gaussian wavepacket. 
This derivation is valid for the case when the two masses of the bipartite system are the same with $m_1 = m_2 = m$. 

\subsection{Preliminaries}
We choose to set the centre-of-mass wavepacket $\varphi(\vec{R})$ to be a Gaussian wavepacket centred at the origin. It is given by 
\begin{equation} \label{app:eq:Gaussian:wavepacket}
\varphi( \vec{R} ) = \frac{1}{\pi^{3/4} \, b^{3/2}} e^{- \vec{R} \cdot \vec{R}/(2b^2)}. 
\end{equation}
Then the full state is given by
\begin{equation} \label{app:eq:initial:state}
\Psi( \vec{r}, \vec{R}) = \psi_{nlm}( \vec{r})   \frac{1}{\pi^{3/4} \,  b^{3/2}} e^{- \vec{R} \cdot \vec{R}/(2b^2)}, 
\end{equation}
where $\psi_{nlm}( \vec{r})$ are the Hydrogenic eigenstates shown in Eq.~\eqref{app:eq:def:Hydrogenic:wavefunctions}. 
We  proceed to compute all the elements of the covariance matrix. Given the definitions of the relative and centre-of-mass coordinates in  Eqs.~\eqref{app:eq:redefined:position:coordinates} and~\eqref{app:eq:redefined:momentum:coordinates}, full two-mode covariance matrix is a  diagonal $12\times12$ matrix that is explicitly given by 
\begin{align} \label{app:eq:CM:elements}
&\sigma_{11} = 2 \braket{x^2} - 2\braket{x } ^2, &&\sigma_{77} = 2\braket{X^2} - 2\braket{X}^2, \nonumber\\
&\sigma_{22} = 2\braket{p_x^2} - 2\braket{p_x}^2, &&\sigma_{88} =2 \braket{P_X^2} -2 \braket{P_X}^2,\nonumber\\
&\sigma_{33} = 2 \braket{y^2} - 2 \braket{y}^2, &&\sigma_{99} = 2\braket{Y^2} - 2\braket{Y}^2, \nonumber\\
&\sigma_{44} = 2\braket{p_y^2} - 2\braket{p_y}^2, &&\sigma_{10,10} =2 \braket{P_Y^2} -2 \braket{P_Y}^2,\nonumber\\
&\sigma_{55} = 2 \braket{z^2} - 2 \braket{z}^2 , &&\sigma_{11,11} = 2\braket{Z^2} - 2\braket{Z}^2, \nonumber\\
&\sigma_{66} = 2\braket{p_z^2} - 2\braket{p_z}^2, &&\sigma_{12,12} =2 \braket{P_Z^2} -2 \braket{P_Z}^2. 
\end{align}
The off-diagonal elements are given by a similar algorithm
\begin{align}
\sigma_{i,j} &=  \braket{\mathbb{X}_i \mathbb{X}_j +\mathbb{X}_j \mathbb{X}_i} -2 \braket{\mathbb{X}_i} \braket{\mathbb{X}_j} ,
\end{align}
for $i \neq j$. However, since the  wavefunction is a product of terms that depend separately on each such variable, we only have to compute the diagonal elements. 
The advantage from working in this basis now becomes evident. Depending on whether the variable of interest belongs to the relative or the centre-of-mass coordinates, 
we find that the other part of the state does not influence the result. In other words, if we are computing a moment of one of the centre-of-mas variables $R_j$, then we find
\begin{equation}
\braket{R_j} =  \int_{\mathbb{R}^6} \mathrm{d} \mathbb{R}^{6} \,  \Psi^*(\vec{r}, \vec{R}) \, R_j \, \Psi( \vec{r}, \vec{R}),  \end{equation}
where $\mathbb{R}^6$ denotes the integration over the full spatial domain of the bipartite state. 
When the state separates in terms of the relative and centre-of-mass coordinates as $\Psi( \vec{r}, \vec{R}) =  \psi_{nlm} ( \vec{r}) \varphi( \vec{R}) $, the part of the state that does not relate to the expectation value in question simply satisfies the normalisation  relation. For example, if we wish to compute the expectation value 
\begin{equation}
\braket{R_j}= \int \mathrm{d}x \mathrm{d} y \mathrm{d}z \, |\psi_{ nlm}(\vec{r)}|^2 \, \int \mathrm{d}X \mathrm{d}Y \mathrm{d}Z \varphi^*( \vec{R}) \, R_j \, \varphi( \vec{R}) \, ,
\end{equation}
where the first integral is just the normalisation condition for $\psi_{nlm}( \vec{r})$, we are left with the task to evaluate the  Gaussian integral. Finally, we remark that since the transformation $S$, shown in Eq.~\eqref{app:eq:symplectic:transformation}, between the $\{\vec{r}_1, \vec{r}_2\}$ basis and the $\{\vec{r}, \vec{R}\}$ basis is symplectic, the Jacobian for the substitution of variables in the integrals is equal to unity.

\subsection{Expectation values and variances for  Gaussian centre-of-mass wavepacket}
If the Gaussian wavepacket is centred at the origin, its expectation values are zero. This is the case for the state we consider,  which is shown in Eq.~\eqref{app:eq:Gaussian:wavepacket}, and thus we find
\begin{equation}
\braket{X} = \braket{Y} = \braket{Z} = 0. 
\end{equation}
The variances, on the other hand, are given by 
\begin{align}
\braket{X^2} &= \frac{1}{ \pi ^{3/2}b^3} \int^\infty_{-\infty} \mathrm{d}X \, \int_{-\infty}^\infty\mathrm{d} Y \, \int_{-\infty}^\infty\mathrm{d}Z \, X^2 \, e^{- (X^2 + Y^2 + Z^2)/b^2} \nonumber  \\
&= \frac{1}{\pi ^{3/2}b^3} \int_{-\infty}^\infty \mathrm{d} Y \, \int_{-\infty}^\infty\mathrm{d}Z e^{- (Y^2 + Z^2)/b^2} \, \int_{-\infty}^\infty \mathrm{d} X \, X^2 \, e^{- X^2/b^2}. 
\end{align}
The first integral is equal to 
\begin{equation} \label{app:eq:Y:Z:integral}
 \int_{-\infty}^\infty \mathrm{d} Y \, \int^\infty_{-\infty}\mathrm{d}Z e^{- (Y^2 + Z^2)/b^2} = b^2 \pi, 
 \end{equation}
and the second integral is equal to 
\begin{equation}
\int_{-\infty}^\infty \mathrm{d} X \, X^2 \, e^{- 2X^2/b^2} = \frac{\sqrt{\pi } b^3}{2}.
\end{equation}
So we are left with 
\begin{align}
\braket{X^2} =\frac{1}{\pi ^{3/2}b^3}b^2 \pi  \frac{\sqrt{\pi} }{2} b^{3}= \frac{b^2}{2}. 
 \end{align} 
This element will be the same for all other quantities. In summary, we have
\begin{align}
\braket{X^2} - \braket{X}^2 &=   \frac{b^2}{2},  \nonumber\\
\braket{Y^2} - \braket{Y}^2 &=  \frac{b^2}{2}, \nonumber\\
\braket{Z^2} - \braket{Z}^2 &=   \frac{b^2}{2}.
\end{align}
Next, we wish to compute the centre-of-mass momentum expectation values. The momentum variable in the $X$ direction is given by 
\begin{equation}
P_X = - i \hbar \frac{d}{dX}, 
\end{equation}
and similarly for $P_Y$ and $P_Z$. Taking the expectation value, we find
\begin{align}
\braket{P_X} &= \frac{1}{ \pi ^{3/2}b^3} \int^\infty_{-\infty} \mathrm{d} X \,e^{-(X^2 + Y^2+ Z^2)/(2b^2)} \left( - i \hbar \frac{d}{dX} \right) e^{- (X^2 + Y^2 + Z^2)/(2b^2)} \nonumber\\
&=  - i \hbar \frac{1}{\pi ^{3/2}b^3}  \int^\infty_{-\infty} \mathrm{d} X \,e^{-(X^2 + Y^2+ Z^2)/(2b^2)}  \left(- \frac{ X}{b^2} \right) e^{- (X^2 + Y^2 + Z^2)/(2b^2)} \nonumber\\
&= i \hbar \frac{1}{\pi ^{3/2} b^5} \int_{-\infty}^\infty \mathrm{d}X \, X \,  e^{- (X^2 + Y^2 + Z^2)/b^2}  = 0, 
\end{align}
where the integral vanishes because the function is odd. Due to the symmetry of the wavepacket, it follows all the expectation values of the momentum coordinates are zero. 

We proceed to compute the variance of the momentum variable. We find
\begin{align}
\braket{\hat{P}_X^2} &=\frac{1}{\pi ^{3/2}b^3}   \int_{-\infty}^\infty \mathrm{d}X \,  \int_{-\infty}^\infty\mathrm{d} Y \,  \int_{-\infty}^\infty\mathrm{d}Z \, e^{- (X^2 + Y^2 + Z^2)/(2b^2)} \left( - i \hbar \frac{d}{dX}\right)^2 e^{- (X^2 + Y^2 + Z^2)/(2b^2)} \nonumber\\
&= -\frac{\hbar^2}{\pi ^{3/2}b^3}  \int_{-\infty}^\infty \mathrm{d} Y \int_{-\infty}^\infty\mathrm{d}Z  \, e^{-  (Y^2 + Z^2)/b^2}  \int_{-\infty}^\infty \mathrm{d}X \,  \left( \frac{X^2}{b^4} - \frac{1}{b^2} \right) e^{- X^2 /b^2}. 
\end{align}
The first  two integrals over $Y$ and $Z$ again evaluate to that in Eq.~\eqref{app:eq:Y:Z:integral}. The second integral becomes
\begin{equation}
\int_{-\infty}^\infty \mathrm{d}X \,  \left( \frac{X^2}{b^4} - \frac{1}{b^2} \right) e^{- X^2 /b^2} = -\frac{1}{2} \sqrt{\pi } \sqrt{\frac{1}{b^2}}. 
\end{equation}	
Putting everything together, we find
\begin{equation}
\braket{P_X^2} = \frac{\hbar^2}{\pi ^{3/2}b^3}\frac{1}{2} \sqrt{\pi } \sqrt{\frac{1}{b^2}}b^2 \pi = \frac{\hbar^2}{2b^2}.
\end{equation}
Again, due to symmetry, we find $\braket{P_X^2} = \braket{P_Y^2} = \braket{P_Z^2}$. So in summary, the momentum variances become
\begin{align}
\braket{P_X^2} - \braket{P_X}^2 &= \frac{\hbar^2}{2  b^2}, \nonumber\\
\braket{P_Y^2} - \braket{P_Y}^2 &= \frac{\hbar^2}{2  b^2},\nonumber \\
\braket{P_Z^2} - \braket{P_Z}^2 &= \frac{\hbar^2}{2  b^2}.
\end{align}
We proceed to consider the Hydrogenic wavefunctions. 

\subsection{Position expectation values of the Hydrogenic subsystems} \label{sec:position:expectation:values:Hydrogen}
We wish to compute the expectation values $
\braket{x}, \braket{y}$,  and $\braket{z}$.
We recall that the Hydrogenic wavefunctions are given by
\begin{equation}
\psi_{nlm}(\vec{r})  = \sqrt{\left( \frac{2}{na_0} \right)^3 \frac{(n-l-1)!}{2n[(n+l)!]^3}} \, e^{- r/na_0} \left( \frac{2r}{na_0} \right)^l \left[ L^{2l+1}_{n-l-1} (2r/na_0) \right] Y^m_l(\theta, \phi) .
\end{equation}
We can rewrite   $\psi_{nlm}(\vec{r})$ in terms of a radial and an angular part
\begin{equation}
\psi_{nlm}( \vec{r}) =  R_{nl}(r) \, Y^m_l (\theta, \phi), 
\end{equation}
 where we have defined 
\begin{align}
R_{nl}(r) = \sqrt{\left( \frac{2}{na_0} \right)^3 \frac{(n-l-1)!}{2n[(n+l)!]^3}}  \,e^{- r/na_0} \left( \frac{2r}{na_0} \right)^l \left[ L^{2l+1}_{n-l-1} (2r/na_0) \right].
\end{align}
In what follows, we will often make use of the absolute value $|Y^m_l(\theta, \phi)|^2$, which is  independent of the variable $\phi$.  To emphasise this fact, we write this as
\begin{equation}
|Y_l^m(\theta, \phi)|^2 = \mathcal{Y}^m_l(\theta).
\end{equation}
We also note that many of the expressions that we wish to compute are easier to evaluate in spherical polar coordinates. The transformation from Cartesian coordinates to spherical polar coordinates is given by
\begin{align} \label{app:eq:cartesians:to:polars}
x &= r \sin \theta \cos \phi ,\nonumber \\
y &= r \sin \theta \sin \phi ,\nonumber \\
z &= r \cos \theta, 
\end{align}
where the infinitesimal volume element transforms as $\mathrm{d}x \, \mathrm{d} y\,\mathrm{d} z = r^2\,\sin\theta  \, \mathrm{d} r \, \mathrm{d}\theta\, \mathrm{d}\phi$. 

We proceed to compute the expectation values of the position  variables $x$, $y$, and $z$.  Starting with $\braket{x}$, the expression that we must calculate is given by
\begin{align}
\braket{x} &= \int_{-\infty}^\infty \mathrm{d} X \, \int_{-\infty}^\infty\mathrm{d}Y \,\int_{-\infty}^\infty \mathrm{d}Z \,  |\varphi(\vec{R})|^2 \int_{-\infty}^\infty \mathrm{d}x \,\int_{-\infty}^\infty \mathrm{d}y \,\int_{-\infty}^\infty \mathrm{d}z \, \psi_{nlm}^*(\vec{r}) \, x \, \psi_{nlm}(\vec{r}) \nonumber \\
&= \int_{-\infty}^\infty \mathrm{d}x \, \int_{-\infty}^\infty\mathrm{d}y \, \int_{-\infty}^\infty\mathrm{d}z \, \psi_{nlm}^*(\vec{r}) \, x \, \psi_{nlm}(\vec{r}). 
\end{align}
Then, using the fact that $x = r \sin \theta  \cos \phi$, we write
\begin{align}
\braket{x} &= \int^\infty_0 \mathrm{d} r\,r^2 \int^\pi_0 \mathrm{d}\theta \, \sin\theta \, \int^{2\pi}_0 \mathrm{d}\phi  \, \psi_{nlm}^*(\vec{r}) \, r \sin\theta \cos\phi \, \psi_{nlm}(\vec{r}) \, \nonumber \\
&= \int^\infty_0 \mathrm{d} r\, r^3 \, R_{nl}^2(r) \,   \int^\pi_0 \mathrm{d}\theta \,  \sin^2\theta\,\int^{2\pi}_0 \mathrm{d}\phi  \,  \cos\phi  \, |Y^m_l (\theta, \phi)|^2\,\nonumber \\
&=  \int^\infty_0 \mathrm{d} r\,r^3 \,   R_{nl}^2(r)  \,   \int^\pi_0 \mathrm{d}\theta \,  \sin^2\theta  \, \mathcal{Y}^m_l(\theta)\,\int^{2\pi}_0 \mathrm{d}\phi \,   \cos\phi \nonumber \\
&= 0, 
\end{align}
which follows because the final integral over $\cos\phi$ is zero. 
Next,  we evaluate $\braket{y}$, where we can use the same trick and write $y =  r \sin \theta \sin \phi$. We have
\begin{align}
\braket{y} &= \int^\infty_{-\infty} \mathrm{d} X \,\int^\infty_{-\infty} \mathrm{d}Y \, \int^\infty_{-\infty}\mathrm{d}Z \,  |\varphi(\vec{R})|^2 \int^\infty_{-\infty} \mathrm{d}x \, \int^\infty_{-\infty}\mathrm{d}y \, \int^\infty_{-\infty}\mathrm{d}z \, \psi_{nlm}^*(\vec{r}) \, y \, \psi_{nlm}( \vec{r}) \, \nonumber \\
&= \int^\infty_{-\infty} \mathrm{d}x \,\int^\infty_{-\infty} \mathrm{d}y \, \int^\infty_{-\infty}\mathrm{d}z \, \psi_{nlm}^*(\vec{r}) \, y \, \psi_{nlm}(\vec{r}) \, \nonumber \\
&= \int^\infty_0 \mathrm{d}r \, r^2 \,  \int^\pi_0  \mathrm{d}\theta \,  \sin\theta \int^{2\pi}_0 \mathrm{d}\phi \, \psi_{nlm}^*(\vec{r}) \, r \,  \sin\theta \,  \sin\phi \, \psi_{nlm}(\vec{r}) \nonumber \\
&= \int_0^\infty\mathrm{d}r \,  r^3  \,  R_{nl}^2(r) \int^\pi_0 \mathrm{d}\theta \, \sin^2\theta \, \mathcal{Y}^m_l(\theta) \,   \int^{2\pi}_0\mathrm{d}\phi \,  \sin\phi  \nonumber \\
&= 0 , 
\end{align}
again, because the last integral over $\phi$ is zero.  Finally, the expectation value $\braket{z}$, with $z = r \cos \theta$, becomes

\begin{align}
\braket{z} &= \int^\infty_{-\infty} \mathrm{d} X \,\int^\infty_{-\infty} \mathrm{d}Y \,\int^\infty_{-\infty} \mathrm{d}Z  \, |\varphi(\vec{R})|^2 \int^\infty_{-\infty} \mathrm{d}x \, \int^\infty_{-\infty}\mathrm{d}y \, \int^\infty_{-\infty}\mathrm{d}z \, \psi_{nlm}^*(\vec{r}) \, z \, \psi_{nlm}(\vec{r}) \, \nonumber \\
&= \int^\infty_{-\infty} \mathrm{d}x \,\int^\infty_{-\infty} \mathrm{d}y \,\int^\infty_{-\infty} \mathrm{d}z \, \psi_{nlm}^*(\vec{r}) \, z \, \psi_{nlm}(\vec{r}) \, \nonumber \\
&= \int^\infty_0 \mathrm{d}r\,   r^2  \, \int^\pi_0 \mathrm{d}\theta \,   \sin\theta \,   \, \int^{2\pi}_0 \mathrm{d}\phi  \, \psi_{nlm}^*( \vec{r}) \, r\, \cos\theta \, \psi_{nlm} (\vec{r})\nonumber \\
&= \int^\infty_0 \mathrm{d}r\,  r^3  \,  R_{nl}^2(r)  \, \int^\pi_0 \mathrm{d}\theta \,   \sin\theta \, \cos\theta \,  \mathcal{Y}_l^m(\theta)  \, \int^{2\pi}_0 \mathrm{d}\phi \nonumber \\
&= 2\pi  \int^\infty_0 \mathrm{d}r \,  r^3  \,  R_{nl}^2(r) \, \int^\pi_0\mathrm{d}\theta \,  \sin\theta \, \cos\theta \,  \mathcal{Y}_l^m(\theta). 
\end{align}
This  angular integral is less trivial.  We focus  our attention on the integral over $\theta$. Given the expression for the spherical harmonics in Eq.~\eqref{app:eq:spherical:harmonics:definition}, we write 
\begin{align}
 \int^\pi_0 \sin\theta \, \cos\theta \,  \mathrm{d}\theta \, \mathcal{Y}_l^m(\theta)  = \frac{2 l + 1}{4\pi} \frac{(l-m)!}{(l+m)!}  \int^\pi_0 \mathrm{d} \theta \, \sin{\theta} \cos{\theta} P^m_l (\cos{\theta}) P^m_l(\cos{\theta}) .
\end{align}
We then perform the substitution $\cos{\theta} = u$. Thus $\mathrm{d}u =- \sin{\theta} \mathrm{d}\theta$, leading to $\mathrm{d}\theta = -\mathrm{d}u/\sin{\theta}$. The limits $(0, \pi)$ become $(1, -1)$, which yields the following integral
\begin{align} \label{app:eq:exp:z:integral}
\int^\pi_0 \mathrm{d} \theta \, \sin{\theta} \cos{\theta} P^m_l (\cos{\theta}) P^m_l(\cos{\theta})  =-  \int^{-1}_1 \mathrm{d} u \,  u \, P^m_l (u) P^m_l(u)  = \int^{1}_{-1} \mathrm{d} u \,  u \, P^m_l (u) P^m_l(u) .
\end{align}
We can now use the following recurrence relation for the associated Legendre polynomials:
\begin{equation} \label{app:eq:associated:legendre:recurrence}
 x P^m_l(x) = \frac{1}{(2 l +1)} \left(  (l - m + 1) P^m_{l+1}(x) + (l+m) P^m_{l-1}(x) \right).
\end{equation}
Inserting this into Eq.~\eqref{app:eq:exp:z:integral} yields
\begin{align} \label{app:eq:Legendre:integral}
\int^\pi_0\mathrm{d}\theta &\, \sin\theta \, \cos\theta \,   \mathcal{Y}_l^m(\theta)  \\
& = \frac{2 l + 1}{4\pi} \frac{(l-m)!}{(l+m)!}  \frac{1}{(2 l +1)}\int^1_{-1} \mathrm{d} u \,   \left(  (l - m + 1) P^m_{l+1}(u) + (l+m) P^m_{l-1}(u) \right) P^m_l(u).\nonumber
\end{align}
However, the orthogonality relation for the associated Legendre polynomials states that 
\begin{equation}
\int^1_{-1} \mathrm{d}u\,  P^m_l(u) P^m_{l'}(u) = \frac{2 (l+1)!}{(2l+1)(l-m)!} \delta_{ll'}, 
\end{equation}
which means that the indices must match in order for the integral in Eq.~\eqref{app:eq:Legendre:integral} to be non-zero. In our case, we have differing indices, and  thus the integral is zero.  This same result  is also noted in Ref.~\cite{samaddar1974some}. 
Therefore, in summary, 
\begin{align}
\braket{x} = 
\braket{y} = 
\braket{z} = 0, 
\end{align}
 which is to be expected since the Hydrogenic subsystem is not expected to display any net movement in any spatial direction.
We proceed with calculating the variances of the  Hydrogenic subsystem.

\subsection{Variances of the Hydrogenic subsystems}
Our goal is to compute the following three variances:
\begin{equation}
\braket{x^2}, \braket{y^2}, \mbox{  and  } \braket{z^2}. 
\end{equation}
To do so, we first present some preliminary  results that will aid our calculations. 

\subsubsection{Preliminaries}
We begin  by introducing the Kramer--Pasternack relation, which provides a closed-form expression for any powers of the expectation value of $\braket{r^q}$.  In general, the expectation value of the variable $r^q$ is given by
\begin{equation}
\braket{r^q} = \int^\infty_0  \mathrm{d}r  \,r^2 \,  (r)^q  R_{nl}^2(r) , 
\end{equation} 
where $q$ is an integer number,  and where $R_{nl}^2(r)$ is the radial wavefunction. 
The Kramer--Pasternack relation reads~\cite{pasternack1937mean,kramers1939quantentheorie}
\begin{equation}
4 ( q+1) \braket{r^q} - 4n^2 (2q+ 1) \braket{r^{q-1}} + n^2 q [(2l+1)^2 - q^2] \braket{r^{q-2}} = 0.
\end{equation}
As will become clear below, we are interested in the second order expression obtained through the Kramer--Pasternack relation. It is given by 
\begin{equation} \label{app:eq:second:order:expectation:values:r}
\bra{\psi_{nlm}}  r^2 \ket{\psi_{nlm}} =  a_0^2 \frac{n^2 (5n^2 - 3l(l+1) + 1)}{2} \, .
\end{equation}
We now proceed to compute the variances  of $x$, $y$, and $z$ one by one. In doing so, we will again make frequent use of the transformation from Cartesian coordinates to spherical polar coordinates shown in Eq.~\eqref{app:eq:cartesians:to:polars}. 

Before we proceed, we also wish to evaluate the following integral, which  will appear a number of times in the calculations below:
\begin{align}
\braket{ \sin^2 \theta \, \mathcal{Y}_l^m(\theta)} = \int^\pi_0 \mathrm{d} \theta \, \sin^3\theta  \, \mathcal{Y}_l^m(\theta) .
\end{align}
 Note the extra factor of $\sin \theta$ in the integral, which arises due to the inclusion of the Jacobian. Using the explicit expressions for the spherical harmonics in Eq.~\eqref{app:eq:spherical:harmonics:definition}, we write:
\begin{align}
\braket{ \sin^2\theta \, \mathcal{Y}_l^m(\theta)} = \frac{2 l + 1}{4\pi} \frac{(l-m)!}{(l+m)!}  \int^\pi_0 \mathrm{d} \theta \, \sin^3{\theta} \,  P^m_l (\cos{\theta}) P^m_l(\cos{\theta}). 
\end{align}
As we did above, we now let $u = \cos{\theta}$, so that $\mathrm{d}\theta = -\mathrm{d}u/\sin{\theta}$. We obtain
\begin{align}
\braket{ \sin^2\theta \, \mathcal{Y}_l^m(\theta)} &=- \frac{2 l + 1}{4\pi} \frac{(l-m)!}{(l+m)!}  \int_1^{-1} \mathrm{d} u \, (1 - u^2) P^m_l (u) P^m_l(u)  \nonumber \\
&= \frac{2 l + 1}{4\pi} \frac{(l-m)!}{(l+m)!}  \int_{-1}^{1} \mathrm{d} u \, (1 - u^2) P^m_l (u) P^m_l(u).
\end{align}
This integral can be divided into two parts: 
\begin{align}
\braket{ \sin^2\theta \, \mathcal{Y}_l^m(\theta)}
&= \frac{2 l + 1}{4\pi} \frac{(l-m)!}{(l+m)!}  \left( \int^1_{-1} \mathrm{d}u \, P_l^m(u) P_l^m(u) - \int^1_{-1} \mathrm{d}u \, u^2 \,P_l^m(u) P_l^m(u)  \right). 
\end{align}
The first integral satisfies the orthogonality relation for the associated Legendre polynomials, which reads
\begin{equation} \label{eq:laguerre:orthogonality:relation}
\int^1_{-1} \mathrm{d}u \, P_l^m(u) P_l^m(u) = \frac{2 (l+m)!}{(2l+1)(l-m)!}, 
\end{equation}
so that we are left with 
\begin{align} \label{app:eq:variance:x2:integral}
\braket{ \sin^2\theta \, \mathcal{Y}_l^m(\theta)} &= \frac{2 l + 1}{4\pi} \frac{(l-m)!}{(l+m)!}  \left(  \frac{2 (l+m)!}{(2l+1)(l-m)!} - \int^1_{-1} \mathrm{d}u \, u^2 \,P_l^m(u) P_l^m(u)  \right) \nonumber \\
&= \frac{1}{2\pi} -  \frac{2 l + 1}{4\pi} \frac{(l-m)!}{(l+m)!} \int^1_{-1} \mathrm{d}u \, u^2 \,P_l^m(u) P_l^m(u) .
 \end{align}
We now use the recurrence relation in Eq.~\eqref{app:eq:associated:legendre:recurrence} to write the remaining integral in Eq.~\eqref{app:eq:variance:x2:integral} as 
\begin{align}
\int^1_{-1} \mathrm{d}u \, u^2 \,P_l^m(u) P_l^m(u)  &= \int^1_{-1} \mathrm{d}u \, \left( \frac{l - m + 1}{2l + 1} \, P^m_{l+ 1}(u) + \frac{l+m}{2l + 1} P^m_{l-1}(u)\right)  \\
&\quad\quad\quad\quad \times \left( \frac{l - m + 1}{2l + 1} \, P^m_{l+ 1}(u) + \frac{l+m}{2l + 1} P^m_{l-1}(u)\right)\nonumber \\
&=  \int^1_{-1} \mathrm{d}u \,\left[ \left( \frac{l - m + 1}{2l + 1} \right)^2 P_{l+ 1}^m(u) P_{l+1}^m(u) + \left( \frac{l + m}{2l + 1} \right)^2 P_{l-1}^m(u) P_{l-1}^m(u) \right]\nonumber, 
 \end{align}
where the cross-terms vanish because of the orthogonality relation in Eq.~\eqref{eq:laguerre:orthogonality:relation}. Then, we use the same relation to find that 
\begin{align} \label{app:eq:legendre:u2}
\int^1_{-1} \mathrm{d}u& \,\left[ \left( \frac{l - m + 1}{2l + 1} \right)^2 P_{l+ 1}^m(u) P_{l+1}^m(u) + \left( \frac{l + m}{2l + 1} \right)^2 P_{l-1}^m(u) P_{l-1}^m(u) \right]  \\
&= \left( \frac{l - m + 1}{2l + 1} \right)^2 \frac{2 (l+1+m)!}{(2(l+1)+1)(l+1-m)!} 
 +  \left( \frac{l + m}{2l + 1} \right)^2\frac{2 (l-1+m)!}{(2(l-1)+1)(l-1-m)!} .\nonumber
\end{align}
Simplifying this expression and multiplying it by the prefactor of the integral, we find
\begin{align}
&\braket{ \sin^2 \theta\, \mathcal{Y}_l^m(\theta)} \nonumber \\
&\quad =  \frac{2 l + 1}{4\pi} \frac{(l-m)!}{(l+m)!}\biggl[  \left( \frac{l  + 1- m}{2l + 1} \right)^2 \frac{2 (l+1+m)!}{(2l+3)(l+1-m)!} +  \left( \frac{l + m}{2l + 1} \right)^2\frac{2 (l-1+m)!}{(2l-1)(l-1-m)!} \biggr] \nonumber \\
&\quad = \frac{1}{4\pi (2l + 1) } \biggl[ \frac{(l-m)!}{(l+m)!} \left( l  + 1- m\right)^2 \frac{2 (l+1+m)!}{(2l+3)(l+1-m)!} \nonumber \\
&\quad \quad \quad \quad \quad \quad \quad \quad+  \frac{(l-m)!}{(l+m)!} \left( l + m \right)^2\frac{2 (l-1+m)!}{(2l-1)(l-1-m)!} \biggr].
\end{align}
Now rename $p = l + m$ and $q = l- m$ to find 
\begin{align}
\braket{ \sin^2\theta \, \mathcal{Y}_l^m(\theta)} &= \frac{1}{4\pi (2l + 1) } \left[ \frac{q!}{p!} \left( q  + 1\right)^2 \frac{2 (p+1)!}{(2l+3)(q+1)!} +  \frac{q!}{p!} \left(p \right)^2\frac{2 (p-1)!}{(2l-1)(q-1)!} \right].
 \end{align}
We then note the following simplifications:
\begin{align}
\frac{(p+1)!}{p!} &= p + 1, &&\frac{p!}{(p+1)!} = \frac{1}{p+ 1},  \nonumber \\
\frac{(p-1)!}{p!} &= \frac{1}{p},  &&
\frac{p!}{(p-1)!} = p \, .
\end{align}
The above expression becomes 
\begin{align}
\braket{ \sin^2\theta \, \mathcal{Y}_l^m(\theta)} &= \frac{1}{4\pi (2l + 1) } \left[ \frac{q!}{p!} \left( q  + 1\right)^2 \frac{2 (p+1)!}{(2l+3)(q+1)!} +  \frac{q!}{p!} \left(p \right)^2\frac{2 (p-1)!}{(2l-1)(q-1)!} \right] \nonumber \\
 &= \frac{1}{4\pi (2l + 1) } \left[ \frac{1}{q+1} (p+1) \frac{2}{2 l + 3} (q + 1)^2 + p^2 q  \frac{1}{p} \frac{2}{2l - 1} \right] \nonumber \\
 &=  \frac{1}{2\pi (2l + 1) } \left[(p+1)(q + 1)  \frac{1}{2 l + 3} + p q   \frac{1}{2l - 1} \right].
 \end{align}
Inserting the original expressions  $p = l + m$ and $q = l - m$, we simplify and find that
 \begin{equation} \label{eq:exp:values:sin3theta:Y}
\braket{  \sin^2\theta \, \mathcal{Y}_l^m(\theta)} =  \frac{l^2+l+m^2-1}{\pi  (2 l-1) (2 l+3)}. 
 \end{equation}
With this relation and the Kramer--Pasternack relation for $\braket{r^2}$ in Eq.~\eqref{app:eq:second:order:expectation:values:r}, we are ready to compute the remaining variances for the Hydrogenic wavefunctions. 

\subsubsection{Calculating $\braket{x^2} $}

The integral for this expectation value is given by 
\begin{align}
\braket{x^2} &= \int^\infty_{-\infty} \mathrm{d}x  \, \int^\infty_{-\infty}\mathrm{d} y \,\int^\infty_{-\infty} \mathrm{d} z \, x^2  \, |\psi_{nlm}(\vec{r})|^2. 
\end{align}
 Again splitting the wavefunction into a radial part $ R_{nl}^2(r)$ and an angular part $|Y^m_l(\theta, \phi)|^2 = \mathcal{Y}^m_l (\theta)$,  and using the fact that $x^2 = r^2 \, \sin^2 \theta \, \cos^2 \phi$, we find
\begin{align}
\braket{x^2} &= \int^\infty_0  \mathrm{d}r \,r^4 \,  R_{nl}^2(r)  \, \int^\pi_0\mathrm{d}\theta \, \sin^3\theta  \,  \mathcal{Y}_l^m(\theta) \int^{2\pi}_0\mathrm{d}\phi \, \cos^2\phi \,   \nonumber \\
&= \pi \int^\infty_0  \mathrm{d}r \,r^4  \,  R_{nl}^2(r)  \int^\pi_0 \mathrm{d}\theta \, \sin^3\theta  \, \mathcal{Y}_l^m(\theta).
\end{align}
We then proceed to use the Kramer--Pasternack relation in Eq.~\eqref{app:eq:second:order:expectation:values:r}  to evaluate the radial integral, and the result for the angular integral  listed in Eq~\eqref{eq:exp:values:sin3theta:Y} to find
\begin{align}
\braket{x^2}
&= a_0^2 \, \frac{n^2 (5n^2 - 3l(l+1) + 1)}{2} \,   \frac{l^2+l+m^2-1}{ (2 l-1) (2 l+3)}.
\end{align}

\subsubsection{Calculating $\braket{y^2} $}
We now proceed with $\braket{y^2 }$,  where we recall that $y^2 = r^2 \, \sin^2 \theta \, \sin^2 \phi$. We find
\begin{align}
\braket{y^2} &= \int^\infty_0  \mathrm{d} r \,r^2 \, \int^\pi_0  \mathrm{d} \theta \, \sin\theta \,  \int^{2\pi}_0 \mathrm{d} \phi  \, y^2  \, |\psi_{nlm}(\vec{r})|^2 \nonumber \\
&= \int^\infty_0  \mathrm{d} r \, r^4 \,  R_{nl}^2(r) \int^\pi_0 \mathrm{d}\theta \, \sin^3\theta \,   \int^{2\pi}_0 \mathrm{d} \phi \,  \sin^2\phi \,  | Y_l^m(\theta, \phi)|^2 .
\end{align}
We start with the last integral,  which, since $|Y_l^m(\theta, \phi)|^2 = \mathcal{Y}_l^m(\theta)$ is independent of $\phi$, becomes
\begin{align}
\int^{2\pi}_0 \mathrm{d}\phi \, \sin^2\phi = \pi. 
\end{align}
Again using the Kramer--Pasternack relation in Eq.~\eqref{app:eq:second:order:expectation:values:r}, and the result for the angular integral in Eq.~\eqref{eq:exp:values:sin3theta:Y}, we find
\begin{equation}
\braket{y^2} = a_0^2 \frac{n^2 (5n^2 - 3l(l+1) + 1)}{2} \frac{l^2+l+m^2-1}{ (2 l-1) (2 l+3)}.
\end{equation}

\subsubsection{Computing $\braket{z^2} $}
This is a bit different, since  $z^2 = r^2 \, \cos^2\theta$. We find
\begin{align}
\braket{z^2} &= \int^\infty_0  \mathrm{d}r \, r^2 \,  R_{nl}^2(r)  \int^\pi_0 \mathrm{d} \theta \, \sin\theta \,   r^2  \,  \cos^2\theta \,   \mathcal{Y}_l^m(\theta) \int^{2\pi}_0 \mathrm{d}\phi \nonumber \\
&= 2\pi \int^\infty_0  \mathrm{d}r \,  r^4 \,  R_{nl}^2(r)  \int^\pi_0 \mathrm{d} \theta \, \sin\theta \, \cos^2\theta \,   \mathcal{Y}_l^m(\theta). 
\end{align}
Starting with the angular integral, and using the expression for the spherical harmonics in Eq.~\eqref{app:eq:spherical:harmonics:definition}, we write
\begin{align}
\int^\pi_0 \mathrm{d}\theta \sin\theta \,  \cos^2\theta \, \mathcal{Y}_l^m(\theta) &= \frac{2 l + 1}{4\pi} \frac{(l-m)!}{(l+m)!}  \int^\pi_0 \mathrm{d} \theta \, \sin{\theta} \cos^2\theta \,  P^m_l (\cos{\theta}) P^m_l(\cos{\theta}). 
\end{align}
We again perform the substitution $u = \cos(\theta)$, and we find $\mathrm{d}u = - \mathrm{d}\theta \sin(\theta)$, so that $\mathrm{d}\theta = - \mathrm{d}u /\sin(\theta)$, which gives
\begin{align}
\int^\pi_0 \mathrm{d}\theta \sin\theta \,  \cos^2\theta \,   P^m_l (\cos{\theta}) P^m_l(\cos{\theta}) &= -\int^{-1}_1 \mathrm{d}u^2 \, u^2 \, P^m_l(u) P^m_l(u) \nonumber \\
&= \int^{1}_{-1} \mathrm{d}u^2 \, u^2 \, P^m_l(u) P^m_l(u).
\end{align}
We already obtained the answer to this quantity in Eq.~\eqref{app:eq:legendre:u2}. The result is
\begin{align}
\frac{2 l + 1}{4\pi}& \frac{(l-m)!}{(l+m)!} \int^{1}_{-1} \mathrm{d}u^2 \, u^2 \, P^m_l(u) P^m_l(u) \nonumber \\
&=  \frac{2 l + 1}{4\pi} \frac{(l-m)!}{(l+m)!} \biggl[ \left( \frac{l - m + 1}{2l + 1} \right)^2 \frac{2 (l+1+m)!}{(2(l+1)+1)(l+1-m)!} 
 \nonumber \\
&\qquad\qquad\qquad\qquad +  \left( \frac{l + m}{2l + 1} \right)^2\frac{2 (l-1+m)!}{(2(l-1)+1)(l-1-m)!} \biggr], 
\end{align}
which can be simplified, so that we ultimately find
\begin{equation} \label{app:eq:sin:cos2:Y:integral}
\int^\pi_0 \mathrm{d}\theta \, \sin\theta \,  \cos^2\theta \,  \mathcal{Y}_l^m(\theta) =\frac{1}{2\pi}\frac{1-2 l^2-2 l+2 m^2}{3 -4 l^2- 4 l}. 
\end{equation}
Then again using the Kramer--Pasternack relation in Eq.~\eqref{app:eq:second:order:expectation:values:r}, we find that 
\begin{equation}
\braket{z^2} = a_0^2 \frac{n^2 (5n^2 - 3l(l+1) + 1)}{2}\frac{1-2 l^2-2 l+2 m^2}{3 -4 l^2- 4 l}. 
\end{equation}
We now have all the covariance matrix elements, and we are ready to compute the entanglement. 

\subsubsection{Summary}
In summary, we have that 
\begin{align}
\braket{x^2} &=a_0^2 \frac{n^2 (5n^2 - 3l(l+1) + 1)}{2} \frac{l^2+l+m^2-1}{ (2 l-1) (2 l+3)} ,  \nonumber \\
\braket{y^2} &=a_0^2 \frac{n^2 (5n^2 - 3l(l+1) + 1)}{2} \frac{l^2+l+m^2-1}{ (2 l-1) (2 l+3)},  \nonumber \\
\braket{z^2} &= a_0^2 \frac{n^2 (5n^2 - 3l(l+1) + 1)}{2}\frac{1-2 l^2-2 l+2 m^2}{3 -4 l^2- 4 l}. 
\end{align}
The fact that $\braket{z^2}$ differs from the other two variances is reasonable given the identification of the rotation axis. 

\subsection{ Expectation values of $p_x$, $p_y$ and $p_z$}
The  expectation values can be quickly computed by realising that we obtain  the same integrals over the spherical harmonics that we already evaluated in Section~\ref{sec:position:expectation:values:Hydrogen}. 

We start with the first  expression $\braket{p_x}$, however we will first compute $\braket{k_x}$ and then multiply by $\hbar$ to obtain the momentum. This quantity is easier to compute in the momentum basis. We therefore use the Fourier transform in Eq.~\eqref{eq:fourier:transform} and Eq.~\eqref{eq:fourier:transform:inverse},  to  write 
\begin{align}
\braket{k_x} &= \int^\infty_0 \mathrm{d}x \int^\infty_0 \mathrm{d}y \int^\infty_0 \mathrm{d}z \, \psi_{nlm}^*(\vec{r}) \left( - i  \frac{\partial}{\partial x} \right) \psi_{nlm}(\vec{r}) \nonumber \\
&=- \frac{i }{(2\pi)^{3/2}}  \int\mathrm{d}\vec{k}'\,  \tilde{\psi}^*( \vec{k}')  \int\mathrm{d}\vec{k} \,  i  k_x  \, \tilde{\psi}( \vec{k}) \int^\infty_0 \mathrm{d}x \,\int^\infty_0 \mathrm{d}y \,\int^\infty_0 \mathrm{d}z \, e^{i (\vec{k} - \vec{k}')\cdot \vec{r}} \nonumber \\
&= - i   \int \mathrm{d}\vec{k}'\, \tilde{\psi}^*( \vec{k}')  \int\mathrm{d}\vec{k} \, i k_x  \,  \tilde{\psi}( \vec{k})  \delta( \vec{k} - \vec{k}') \nonumber \\
&=  \int \mathrm{d}\vec{k}  \, \tilde{\psi}^*( \vec{k}) \,   k_x  \,  \tilde{\psi}( \vec{k}). 
\end{align}
The Hydrogen wavefunction in the momentum representation is given in Eq.~\eqref{app:eq:Hydrogen:wavefunction:momentum:space}  in terms of the wavevector $\vec{k}$. 
Using this expression, and moving to spherical coordinates $(k, \theta, \phi)$, where $k_x = k \cos \theta \sin \phi$, we find
\begin{align}
\braket{k_x} &=  \int^\infty_{- \infty} \mathrm{d} k_x \int^\infty_{- \infty}  \mathrm{d}k_y \int^\infty_{- \infty} \mathrm{d} k_z  \, \tilde{\psi}^*( \vec{k}) \,   k_x  \,  \tilde{\psi}( \vec{k})  \,  \nonumber \\
&= \int^\infty_{0}  \mathrm{d} k \,k^2  \int^\pi_0 \mathrm{d}\theta \, \sin \theta \,\int^{2\pi}_0 \mathrm{d}\phi \, k \,  \cos\theta \,  \sin \phi \,   |F_{nl}(k)|^2 |Y_l^m(\theta, \phi)|^2 \nonumber \\
&=  \int _0^\infty \mathrm{d} k  \, k^3\, |F_{nl}(k)|^2\int^\pi_0 \mathrm{d}\theta \sin \theta   \cos \theta \,    \mathcal{Y}_l^m (\theta)  \, \int^{2\pi}_0 \mathrm{d}\phi \,   \sin \phi , 
\end{align}
where we again used the fact that $|Y_l^m(\theta, \phi)|^2  = \mathcal{Y}_l^m(\theta)$, which is independent of $\phi$. However, we again note that the last integral is again zero, and thus $\braket{p_x} = 0$. In fact, since the angular integral is the same for the position and momentum expectation values, we conclude that we also have $\braket{p_y}= \braket{p_z} = 0$. Intuitively, this is reasonable, since a non-zero momentum expectation value would mean that the system has a net non-zero motion in one of the directions.

\subsection{Variance of $p_x^2$, $p_y^2$ and $p_z^2$}
We proceed to compute the variances of the relative momentum variables. In the wavevector basis, we find that 
\begin{align}
\braket{k_x^2} &= \int^\infty_{-\infty} \mathrm{d} kp_x \int^\infty_{-\infty} \mathrm{d} kp_y \int^\infty_{-\infty} \mathrm{d} k_z \, \tilde{\psi}^* _{nlm} ( \vec{k}) \, k_x^2 \, \tilde{\psi}_{nlm}(\vec{k}) \nonumber \\
&= \int_0^\infty \mathrm{d}p \, p^4  |F_{nl}(p)|^2 \int^\pi_0 \mathrm{d} \theta\,  \sin^3 \theta  \int^{2\pi}_0  \mathrm{d} \phi \,\cos ^2 \phi \,  |Y_{l}^m(\theta, \phi)|^2  \nonumber \\
&= \pi  \int \mathrm{d}p \, p^4  |F_{nl}(p)|^2 \int \mathrm{d} \theta \, \sin^3 \theta \,   \mathcal{Y}_{l}^m(\theta, \phi).
\end{align}
By using the result for the weighted Gegenbauer polynomials in Eq.~\eqref{app:eq:weighted:Gegenbauer:relation}, and the expectation value for the angular integral in Eq.~\eqref{eq:exp:values:sin3theta:Y}, we find that
\begin{align}
\braket{k_x^2} 
&= \left( \frac{1}{a_0 n} \right)^2  \frac{l^2+l+m^2-1}{  (2 l-1) (2 l+3)}. 
\end{align}
By symmetry, and using the results we derived in the previous section, the other values become:
\begin{align}
\braket{k_y^2} &= \left( \frac{1}{a_0 n} \right)^2  \frac{l^2+l+m^2-1}{  (2 l-1) (2 l+3)}, 
\end{align}
and,  using the result in Eq.~\eqref{app:eq:sin:cos2:Y:integral}, we have 
\begin{equation}
\braket{k_z^2} =2 \pi \left( \frac{1}{a_0 n} \right)^2 \braket{ \sin \theta \,  \cos^2 \theta  \,\mathcal{Y}_l^m(\theta) } = \left( \frac{1}{a_0 n} \right)^2\frac{1-2 l^2-2 l+2 m^2}{ 3 - 4 l^2-4 l}. 
\end{equation}

\subsection{Summary of expectation values and variances}
The expectation values and variances for the centre-of-mass variables are
\begin{align}
&\braket{X} = \braket{Y} = \braket{Z} = 0, \nonumber \\
&\braket{X^2} = \braket{Y^2} = \braket{Z^2} = \frac{b^2}{2} , \nonumber \\
&\braket{P_X} = \braket{P_Y} = \braket{P_Z} = 0,\nonumber \\
& \braket{P_X^2} = \braket{P_Y^2} = \braket{P_Z^2} =\frac{\hbar^2}{2 b^2}.
\end{align}
And the expectation values and variances for the relative variables are
\begin{align}
&\braket{x} = \braket{y} = \braket{z} = 0,  \nonumber \\
&\braket{x^2} = \braket{y^2} =a_0^2 \frac{n^2 (5n^2 - 3l(l+1) + 1)}{2} \frac{l^2+l+m^2-1}{ (2 l-1) (2 l+3)}, \nonumber \\
&\braket{z^2} = a_0^2 \frac{n^2 (5n^2 - 3l(l+1) + 1)}{2}\frac{1-2 l^2-2 l+2 m^2}{3 -4 l^2- 4 l},  \nonumber \\
&\braket{p_x} = \braket{p_y} = \braket{p_z} = 0,  \nonumber \\
&\braket{p_x^2} = \braket{p_y^2} = \left( \frac{\hbar}{a_0n} \right)^2  \frac{l^2+l+m^2-1}{  (2 l-1) (2 l+3)},  \nonumber \\
&\braket{p_z^2} =  \left( \frac{\hbar}{a_0 n} \right)^2\frac{1-2 l^2-2 l+2 m^2}{ 3 - 4 l^2-4 l}. 
\end{align}
This ends our investigation into the second moments of a Hydrogenic system.

\end{document}